\documentclass[12pt,reqno]{article}

\usepackage{mathtools}
\usepackage{dirtytalk}
\usepackage{amsmath}
\usepackage{amsfonts}
\usepackage{amssymb}
\usepackage{amsxtra}
\usepackage{amsthm}
\usepackage{graphicx}
\usepackage{caption}
\usepackage{ushort}
\usepackage{color}
\usepackage{hyperref}
\usepackage[parsep]{collref}
\hypersetup{linktocpage=true}
\usepackage{enumitem}
\usepackage{titlesec}
\usepackage{cite}
\usepackage{slashed}
\usepackage{braket}
\usepackage{flexisym}
\usepackage{xcolor}
\usepackage{dsfont}
\usepackage{titlesec}
\usepackage{romannum}
\usepackage{caption}
\usepackage{subcaption}
\usepackage{mathrsfs}
\usepackage[bottom]{footmisc}

\newcommand\numberthis{\addtocounter{equation}{1}\tag{\theequation}}

\newcommand{\mb}{\mathbb}
\newcommand{\mc}{\mathcal}

\newcommand{\ph}{\phantom}

\DeclareMathOperator{\vol}{vol}

\newcommand{\imperial}{\it The Blackett Laboratory, Imperial College London\\
Prince Consort Road, London SW7 2AZ}

\newcommand{\auth}{K. C. Matthew Cheung and Rahim Leung}

\let\oldabstract\abstract
\let\oldendabstract\endabstract
\makeatletter
\renewenvironment{abstract}
{%
               {\list{}{\addtolength{\leftmargin}{1em} 
                        \listparindent 1.5em%
                        \itemindent    \listparindent%
                        \rightmargin   \leftmargin%
                        \parsep        \z@ \@plus\p@}%
                \item\relax}%
               {\endlist}%
\oldabstract}
{\oldendabstract}
\makeatother

\numberwithin{equation}{subsection}

\let\oldsection\section
\renewcommand{\section}{\renewcommand{\theequation}{\thesection.\arabic{equation}}\oldsection}
\let\oldsubsection\subsection
\renewcommand{\subsection}{\renewcommand{\theequation}{\thesubsection.\arabic{equation}}\oldsubsection}

\titlespacing*{\section}
{0pt}{5.5ex plus 1ex minus .2ex}{4.3ex plus .2ex}
\titlespacing*{\subsection}
{0pt}{5.5ex plus 1ex minus .2ex}{4.3ex plus .2ex}

\begin{document}
\setcounter{page}{0}
\thispagestyle{empty}

\begin{center}  

{\Large {\bf Type IIA embeddings of $D=5$ minimal gauged supergravity via Non-Abelian T-duality}}   

\vspace{15pt}

\auth

\vspace{7pt}
\imperial

\end{center} 

\vspace{10pt}

\begin{abstract}


\noindent In this note, we construct explicit Type IIA uplifts of $D=5$ minimal gauged supergravity, by T-dualising known Type IIB uplifts on $N_5 = S^5$, $T^{1,1}$ and $Y^{p,q}$ along their $SU(2)$ isometries. When the $D=5$ gauge field is set to zero, our uplifts recover precisely the known non-Abelian T-duals of the $AdS_5\times N_5$ solutions.  As an application, we obtain new supersymmetric $AdS_3\times\Sigma\times M_5$ solutions in Type IIA, where $\Sigma = \mathbb{WCP}^1_{[n_-,n_+]}$ is a weighted projective space. Existing holographic results of T-dualised AdS solutions suggest that our solutions capture features of $d = 2$ SCFTs with $\mathcal{N}=(0, 2)$ supersymmetry.

\end{abstract}
\vfill\leftline{}\vfill
\pagebreak

\tableofcontents
\addtocontents{toc}{\protect\setcounter{tocdepth}{2}}
\pagenumbering{arabic}
\setcounter{page}{1}
\setcounter{footnote}{0}

\section{Introduction}\label{sec:introduction}


Non-Abelian T-duality (NATD) was first introduced in \cite{delaOssa:1992vci} as the generalisation of the Buscher procedure for Abelian T-duality \cite{Buscher:1987sk,Buscher:1987qj} at the level of the string $\sigma$-model. Unlike its Abelian counterpart, NATD is not established as an exact quantum symmetry of string theory \cite{Giveon:1993ai,Alvarez:1994np,Sfetsos:1994vz,Lozano:1995jx}, mainly due to the two following reasons. The results of \cite{delaOssa:1992vci} are only applicable at tree level in string perturbation theory, and NATD is only known to respect conformal symmetry to first-order in $\alpha'$ corrections. Despite its rather tenuous status upon quantisation, this should not be seen as an obstruction to construct new supergravity solutions using NATD. The work of \cite{delaOssa:1992vci} focused purely on the NSNS sector, and in \cite{Sfetsos:2010uq}, a consistent set of rules on how to implement NATD in backgrounds with non-vanishing RR fields was provided. Since then, NATD has proved to be a powerful technique in generating new, highly non-trivial supergravity solutions with many interesting holographic applications \cite{Lozano:2011kb,Itsios:2012dc,Lozano:2012au,Itsios:2012zv,Itsios:2013wd,Jeong:2013jfc,Sfetsos:2014tza,Kelekci:2014ima,Macpherson:2014eza,Araujo:2015npa,Macpherson:2015tka,PandoZayas:2015auw,Itsios:2017cew,Lozano:2018pcp,Lozano:2019ywa,Lozano:2013oma,Lozano:2016kum}. Notable examples are the $SU(2)$ T-duals of $AdS_5\times S^5$ and $AdS_5\times T^{1,1}$ (i.e. Klebanov-Witten geometry \cite{Klebanov:1998hh}), constructed in \cite{Sfetsos:2010uq} and \cite{Itsios:2012zv,Itsios:2013wd} respectively. Rather remarkably, the $SU(2)$ T-dual of $AdS_5\times S^5$ is shown to capture features of the $\mc{N}=2$ Gaiotto-Maldacena background \cite{Gaiotto:2009gz,Reid-Edwards:2010vpm} dual to $\mc{N}=2$ theories discussed in \cite{Gaiotto:2009we}. Meanwhile, the $SU(2)$ T-dual of $AdS_5\times T^{1,1}$ is proposed as a dual to an $\mc{N}=1$ theory arising as the IR fixed point of M5-branes wrapped on a two-sphere \cite{Bah:2012dg}. One key message from all the examples in \cite{Sfetsos:2010uq,Lozano:2011kb,Itsios:2012dc,Lozano:2012au,Itsios:2012zv,Itsios:2013wd,Jeong:2013jfc,Sfetsos:2014tza,Kelekci:2014ima,Macpherson:2014eza,Araujo:2015npa,Macpherson:2015tka,PandoZayas:2015auw,Itsios:2017cew,Lozano:2018pcp,Lozano:2019ywa,Lozano:2013oma,Lozano:2016kum} is that NATD seems to produce AdS/CFT pairs with very different features from the original, non-dualised ones. 

\begin{figure}[h]
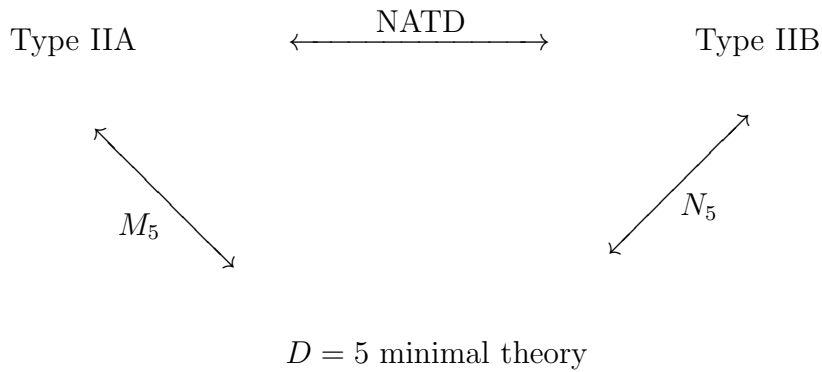

\begin{equation}
    \begin{aligned}
        & \text{Type IIA}
        & \xleftrightarrow[\hspace{3.25cm}]{\text{NATD}}\hspace{0.5cm}
        & \text{\hspace{1.425cm}Type IIB}\\
        & \hspace{5mm}\rotatebox[origin=c]{315}{$\xleftrightarrow[{\rotatebox[origin=c]{45} {$M_5$}}]{\phantom{\text{expansion1234}}}$}
        & \hspace{2.15mm}\resizebox{10mm}{7.5mm}{$\phantom{xx}$}\hspace{4.25mm}
        & \hspace{-1.3mm}\rotatebox[origin=c]{225}{$\xleftrightarrow[\phantom{\text{expansion1234}}]{\rotatebox[origin=c]{135} {$N_5$}}$}\\
        &\hspace{0.0cm}
        & \text{\hspace{-0.1cm}$D=5$ minimal theory}\nonumber
    \end{aligned}
\end{equation}
\caption{A commuting diagram showing the different ways of obtaining/uplifting the minimal gauged supergravity in five dimensions from/to Type II supergravities.}
\label{fig:1}
\end{figure}

Our main goal in this note is to use NATD, a self-contained review of which we will provide in Section \ref{section:NATD basics}, to obtain new embeddings of $D=5$ minimal gauged supergravity theory into Type IIA supergravity.\footnote{Exact $D=10$ embeddings of some lower-dimensional supergravity theories, such as $F(4)$ gauged supergravity in $D=6$, via NATD have been constructed in \cite{Itsios:2012dc,Jeong:2013jfc}.} \footnote{We note that it is also possible to uplift $D=5$ minimal gauged supergravity to massive Type IIA supergravity following the procedure outlined in \cite{Faedo:2019cvr}, which comes from the fact that one can truncate massive Type IIA supergravity to $D=7$ minimal gauged supergravity \cite{Passias:2015gya}, and further truncate the $D=7$ theory on a Riemann surface with topological twist to obtain the minimal gauged supergravity in $D=5$ \cite{Faedo:2019cvr}.} This five-dimensional theory captures some essential features of the worldvolume theory of D3-branes, and can be uplifted to Type IIB supergravity on any five-dimensional Sasaki-Einstein manifold $N_5$ \cite{Buchel:2006gb,Gauntlett:2007ma}. When $N_5$ admits a non-Abelian isometry, it is possible to utilise the technology of NATD to dualise the IIB uplift formulae to Type IIA supergravity. The resulting configuration describes a consistent dimensional reduction of Type IIA supergravity on a new manifold $M_5$ to $D=5$ minimal gauged supergravity. For our purposes, we will be focusing on Sasaki-Einstein manifolds with an $SU(2)$ isometry, specifically $S^5$, $T^{1,1}$ and $Y^{p,q}$ \cite{Gauntlett:2004yd}. These will provide us with three inequivalent embeddings of $D=5$ minimal gauged supergravity into Type IIA supergravity, which we will detail in Section \ref{section:IIAembedding}. The dualisation procedure is summarised in Figure \ref{fig:1}.

With our new embeddings, we will provide a new class of exact T-dual solutions of Type IIA in Section \ref{section:NATD_S5} and discuss their properties and implications.  Specifically, we construct supersymmetric solutions of the form $AdS_3\times \Sigma \times M_5$, where $\Sigma$ is the weighted projective space $\mathbb{WCP}^1_{[n_-,n_+]}$, also known as a spindle. These spindle solutions were first constructed in \cite{Ferrero:2020laf,Ferrero:2020twa}, providing a novel way to wrap branes over orbifolds, where the supersymmetry is not realised via the usual topological twist. Here, it is the $AdS_3\times \Sigma$ solution of \cite{Ferrero:2020laf} that is of interest, which corresponds to the near-horizon limit of D3-branes wrapping on a spindle. These new $AdS_3\times \Sigma \times M_5$ solutions of Type IIA are all supersymmetric, preserving the same amount of supersymmetry as the original solutions. When the original, non-dualised geometry $N_5$ is $T^{1,1}$ or $Y^{p,q}$, these solutions are also smooth and can potentially be viewed as dual to $d=2$ SCFTs which arise from compactifying $d=4$, $\mc{N}=1$ theories in \cite{Bah:2012dg} on a spindle. 
 

We will conclude with a few remarks in Section \ref{conclusion}, and collect some useful results in the appendices, including a corrected version of the Type IIB $S^5$ reduction in the bosonic sector, a derivation of the $S^5$ reduction in the fermionic sector, as well as an alternative reduction of the fermionic sector in the $Y^{p,q}$ case in a frame that is more suited to NATD.


\section{NATD basics}\label{section:NATD basics}

In this section, we will provide a brief, but self-contained, review of the NATD rules for Type II supergravity, mostly following the conventions of \cite{Itsios:2013wd}. We will also focus only on target spaces equipped with an $SU(2)$ isometry. For more general isometry groups, we refer readers to \cite{Lozano:2011kb}.

Consider a Type II (A or B) supergravity background that supports an $SU(2)$ isometry. Starting with the common NSNS sector, the string frame metric has the form\footnote{All Type II metrics presented in this note are in string frame. To transition to Einstein frame, take $ds^2_{Ein} = e^{-\Phi/2}ds^2_{str}$.}
\begin{equation}\label{undualmetric}
ds^2=G_{\mu\nu}(x)dx^\mu dx^\nu+2G_{\mu i}(x)dx^\mu L^i +g_{ij}(x)L^iL^j\,,
\end{equation}
where $\mu,\nu\in\{0,\ldots,6\}$, and we denote by $L=\mathrm{g}^{-1}d\mathrm{g}$, the $SU(2)$ left-invariant Maurer-Cartan one-forms, with $\mathrm{g}\in SU(2)$ given by \eqref{su2groupelement}. The NSNS 2-form is 
\begin{equation}
B_{(2)}=\frac{1}{2}B_{\mu\nu}(x)dx^\mu\wedge dx^\nu+B_{\mu i}(x)dx^\mu\wedge L^i+\frac{1}{2}b_{ij}(x)L^i\wedge L^j\,,
\end{equation}
and the dilaton is given by
\begin{equation}
\Phi=\Phi(x)\,.
\end{equation}
The Lagrangian of the non-linear $\sigma$-model encoding the NSNS sector is
\begin{equation}
\mathcal{L}_{\sigma}=Q_{\mu\nu}\partial_+X^\mu\partial_{-}X^\nu+Q_{\mu i}\partial_+X^\mu L_-^i+Q_{i\mu} L_+^i\partial_-X^\mu+E_{ij}L^i_+L^j_-\,,
\end{equation}
where
\begin{equation}
Q_{\mu\nu}=G_{\mu\nu}+B_{\mu\nu}\,,\quad Q_{\mu i}=G_{\mu i}+B_{\mu i}\,,\quad Q_{i\mu}=G_{i\mu}+B_{i\mu}\,,\quad E_{ij}=g_{ij}+b_{ij}\,.
\end{equation}
The NATD procedure outlined in \cite{Itsios:2013wd,Sfetsos:2010uq,Lozano:2011kb,Kelekci:2014ima} begins by gauging the sigma model with the non-Abelian isometry of the target space, replacing
\begin{equation}
\partial_{\pm}\mathrm{g}\rightarrow D_{\pm}\mathrm{g}=\partial_{\pm}\mathrm{g}-A_{\pm}\mathrm{g}\,.
\end{equation}
Then, the gauge fields $A_{\pm}$ are made non-dynamical by introducing the Lagrange multipliers $v_i$,
\begin{equation}
-i\alpha'\eta\int d^2\sigma\, \text{Tr}\left(vF_{\pm}\right)\quad \text{with}\quad F_{\pm}=\partial_+A_--\partial_-A_+-[A_+,A_-]\,,
\end{equation}
to invoke a flat connection. Here, $\eta=\pm 1$ denotes the right/left-moving frame, and the $\alpha'$ factor is included to ensure the dual coordinates $v_i$ remain dimensionless. For notational convenience, we will now set $\alpha'=1$. Finally, by integrating out the gauge fields, we obtain the dual Lagrangian
\begin{equation}\label{duallag}
\hat{\mathcal{L}}_{\sigma}=Q_{\mu\nu}\partial_+X^\mu\partial_{-}X^\nu+\left[\partial_+v_i+\partial_+X^\mu Q_{\mu i}\right]\left(M^{-1}\right)_{ij}\left[\partial_-v_j-Q_{j\mu}\partial_-X^\mu\right]\,,
\end{equation}
where
\begin{equation}\label{eq:Mij_matrix}
M_{ij}=E_{ij}+\eta f_{ij}\,,\quad \,\text{with}\, \quad  f_{ij}=f_{ij}^{\phantom{ij}k}v_k\,,
\end{equation}
and $f_{ij}^{\ph{ij}k}$ are the $SU(2)$ structure constants. The dual Lagrangian allows us to read off the target space metric and the NSNS 2-form of the T-dual theory as follows
\begin{equation}
\begin{split}
&\hat{Q}_{\mu\nu}=Q_{\mu\nu}-Q_{\mu i}\left(M^{-1}\right)_{ij} Q_{j\nu}\,,\quad \hat{E}_{ij}=\left(M^{-1}\right)_{ij}\,,\\
&\hat{Q}_{\mu i}=Q_{\mu j}\left(M^{-1}\right)_{ji}\,,\quad \hat{Q}_{i\mu}=-\left(M^{-1}\right)_{ij}Q_{j\mu}\,.
\end{split}
\end{equation}
Finally, the dilaton receives a contribution at the quantum level, and it is given by
\begin{equation}
\hat{\Phi}=\Phi-\frac{1}{2}\log\left(\text{det}M\right)\,.
\end{equation}
To compute the transformation of the RR fluxes, we first note that the dualised non-linear sigma model \eqref{duallag} defines two separate frames along the dualised $SU(2)$ directions. More specifically, we find from \eqref{duallag},
\begin{equation}\label{frametransformation}
\begin{split}
&\hat L^i_{+} = -(M^{-1})_{ji}\left(\partial_+ v_j+Q_{\mu j}\partial_+ X^\mu\right) \,,\\
&\hat L^i_{-} = (M^{-1})_{ij}\left(\partial_- v_j-Q_{j\mu}\partial_- X^\mu\right)\,.
\end{split}
\end{equation}
Now, by defining the matrices $\kappa^a_i$ and vectors $\lambda^a_\mu$ to be 
\begin{equation}
g_{ij} =\kappa^a_i\kappa^a_j \,,\quad G_{i\mu} = \kappa^a_i\lambda^a_\mu\,,
\end{equation}
the orthonormal frame along the $SU(2)$ directions associated to the original, non-dualised metric \eqref{undualmetric} is given by
\begin{equation}
e^a = \kappa^a_i L^i + \lambda^a_\mu dX^\mu \,.
\end{equation}
Thus, \eqref{frametransformation} leads to two separate frames
\begin{equation}\label{leftrightvielbeins}
\begin{split}
&\hat e^a_{+} = -\kappa^a_i(M^{-1})_{ji}\left(dv_j + Q_{\mu j}dX^\mu\right) + \lambda^a_\mu dX^\mu \,,\\
&\hat e^a_- = \kappa^a_i(M^{-1})_{ij}\left(dv_j - Q_{j\mu}dX^\mu\right) + \lambda^a_\mu dX^\mu \,.
\end{split}
\end{equation}
We will abbreviate this in matrix notation by 
\begin{equation}
\hat e_+ =-\kappa M^{-T}\left(dv+Q^TdX\right) + \lambda dX\,,\quad \hat e_- =\kappa M^{-1}\left(dv-QdX\right) + \lambda dX \,,
\end{equation}
where $(Q)_{i\mu} = Q_{i\mu}$. These two frames do not describe different physics, since $\hat e^a_+ \hat e^a_+ = \hat e^a_-\hat e^a_-$. As such, there is an orthogonal transformation $\Lambda \in O(3)$ relating the two frames,
\begin{equation}
\hat e_+ = \Lambda \hat e_- \,.
\end{equation}
By equating the $dv$ terms, we find that this orthogonal transformation is given by
\begin{equation}
\Lambda = -\kappa M^{-T}M\kappa^{-1} \,,
\end{equation}
with $\det\Lambda = -1$. The $dX$ terms also equate identically. Defining the flat index coordinate $\zeta^a$ via
\begin{equation}
\left(\kappa^{-T}M\kappa^{-1}\right)_{ab}=\delta_{ab}+\eta\epsilon_{abc}\zeta^c\,,
\end{equation}
the matrix $\Lambda$ can be written as
\begin{equation}
\Lambda_{ab} = \frac{\zeta^2-1}{\zeta^2+1}\delta_{ab} - \frac{2}{\zeta^2+1}\left(\zeta_a\zeta_b +\eta\epsilon_{abc}\zeta_c\right) \,.
\end{equation}
This orthogonal transformation induces an action on spinors through the following transformation of the gamma matrices,
\begin{equation}
\Omega^{-1}\Gamma^a\Omega = \Lambda^a_{\ph{a}b}\Gamma^b \,.
\end{equation}
The solution to this, up to a global sign, is
\begin{equation}
\Omega=\frac{1}{\sqrt{\zeta^2+1}}\left(\Gamma^{789}-\eta\, \zeta_a\Gamma^a\right)\Gamma_{11}\,,
\end{equation}
and its inverse is
\begin{equation}
\Omega^{-1}=\frac{1}{\sqrt{\zeta^2+1}}\left(\Gamma^{789}+\eta\, \zeta_a\Gamma^a\right)\Gamma_{11}\,,
\end{equation}
where $\Gamma_{11} = \Gamma^0\cdots\Gamma^9$ with $\Gamma_{11}^2 = \mathds{1}$. Finally, to calculate the transformation of the RR fluxes, we package them into the bispinors
\begin{equation}
\begin{split}
\text{Type \Romannum{2}B}:\quad P&=\frac{e^\Phi}{2}\sum_{n=0}^4{\slashed{F}}_{(2n+1)}\,,\\
\text{Type \Romannum{2}A}:\quad \hat{P}&=\frac{e^{\hat\Phi}}{2}\sum_{n=0}^5{\slashed{\hat{F}}}_{(2n)}\,,
\end{split}
\end{equation}
where 
\begin{equation}
\slashed{F}_{(n)} = \frac{1}{n!}F_{M_1\cdots M_n}\Gamma^{M_1\cdots M_n} \,,
\end{equation}
and the NATD procedure relates these bispinors via
\begin{equation}\label{eq:RR_transformation_rule}
\hat{P}=P\cdot \Omega^{-1}\,.
\end{equation}
The bispinors $P$ and $\hat P$ are written in the democratic convention, where higher rank forms are related to lower rank ones by Hodge duality. To be precise, we have\footnote{Our convention for the Hodge dual is $\ast(dx^{m_1}\wedge\cdots\wedge dx^{m_p}) = \frac{1}{q!}\sqrt{|g|}\epsilon_{\ph{m_1\cdots m_p}n_1\cdots n_q}^{m_1\cdots m_p} dx^{n_1}\wedge\cdots\wedge dx^{n_q}$, where $\epsilon_{n_1\cdots n_D}$ with lowered indices is numerical, and $q = D-p$.}
\begin{equation}\label{highlowformsdual}
F_{(n)} = (-1)^{\frac{n(n-1)}{2}}{\ast}F_{(10-n)} \,.
\end{equation}
One thing to note is that since $\Lambda\neq\Lambda^{-1}$, the $\Omega^{-1}$ matrix changes when we switch frames (i.e. using $\hat e_-$ as opposed to $\hat e_+$). For concreteness, our convention here will be the same as \cite{Itsios:2013wd}. We will pick $\eta = +1$ and use the right-moving frames $\hat e_+$. The $\Omega^{-1}$ matrix is then
\begin{equation}\label{eq:omega_matrix_we_use}
\Omega^{-1}=\frac{1}{\sqrt{\zeta^2+1}}\left(\Gamma^{789} + \zeta_a\Gamma^a\right)\Gamma_{11}\,.
\end{equation}
A general formula for the transformed RR fluxes is also given in equation (C.11) of \cite{Itsios:2013wd}.

\section{Uplifting $D=5$ minimal gauged supergravity to IIA}\label{section:IIAembedding}

In this section, we will provide new uplift formulae of $D=5$ minimal gauged supergravity by T-dualising the Type IIB uplifts on $S^5$, $T^{1,1}$ and $Y^{p,q}$. The Type IIB uplifts on these manifolds can be found in Appendix \ref{sec:IIB_truncations_on_SE5}. Before presenting our results, below is a lightning review of $D=5$ minimal gauged supergravity.

The bosonic sector of $D=5$ minimal gauged supergravity \cite{Gunaydin:1983bi} contains only the metric $g_{\mu\nu}$ and the graviphoton $A_{(1)}$, and the equations of motion are given by
\begin{equation}
\begin{split}
&R_{\mu\nu} = -4g^2g_{\mu\nu} + \frac{2}{3}{F}^{}_{\mu\rho}{F}_{\nu}^{\ph{\nu}\rho} - \frac{1}{9}g_{\mu\nu}\left({F}_{(2)}\right)^2 \,,\\
&d{\ast_5}{F}_{(2)} = -\frac{2}{3}{F}_{(2)}\wedge{F}_{(2)} \,,
\end{split}
\end{equation}
where ${F}_{(2)} =d{A}_{(1)}$, and $g$ is the gauge coupling. A solution is supersymmetric if it admits a Dirac spinor satisfying the Killing spinor equation
\begin{equation}\label{eq:5d_minimal_KSE}
\left[\nabla_{\alpha}^{(5)} -igA_\alpha+\frac{i}{12}\left(\rho_{\alpha}^{\phantom{\alpha}\beta\gamma}-4\delta_{\alpha}^\beta\rho^\gamma\right)F_{\beta\gamma}+\frac{g}{2}\rho_\alpha\right]\chi=0\,.
\end{equation}
For our purposes, the $\text{Cliff}(4,1)$ gamma matrices are chosen to be
\begin{equation}
\begin{split}
\rho^0&=i\tau_2\otimes \tau_3\,,\quad \rho^1=-\tau_1\otimes \tau_3\,,\quad\rho^2=-\tau_3\otimes \tau_3\,,\\
\rho^3&=-\mathds{1}\otimes \tau_2\,,\quad\rho^4=-\mathds{1}\otimes \tau_1\,,
\end{split}
\end{equation}
where $\tau_1,\tau_2,\tau_3$ are the standard Pauli matrices obeying $\tau_i\tau_j=\delta_{ij}\mathds{1}+i\epsilon_{ijk}\tau_k$. This representation of the gamma matrices satisfies $\{\rho_{\alpha},\rho_{\beta}\}=2\eta_{\alpha\beta}$ and $\rho^{01234}=-i\mathds{1}$. 

Any solution to this five-dimensional theory can be uplifted to Type IIB supergravity, either on $S^5$ via \cite{Cvetic:2000nc} or more generally on a Sasaki-Einstein manifold via \cite{Buchel:2006gb,Gauntlett:2007ma} (see also Appendix \ref{sec:IIB_truncations_on_SE5}). As we mentioned before, the Sasaki-Einstein manifolds of interest are those that possess an $SU(2)$ isometry with which we can T-dualise. The Sasaki-Einstein manifolds considered in this note are $S^5$, $T^{1,1}$ and $Y^{p,q}$. The maximally supersymmetric $AdS_5\times S^5$ vacuum of Type IIB supergravity is the most widely celebrated example of the AdS/CFT correspondence, and is known to be dual to the $\mc{N}=4$ SYM theory in four dimensions \cite{Maldacena:1997re}. Meanwhile, the supersymmetric $AdS_5\times T^{1,1}$ solution, also known as the Klebanov-Witten geometry \cite{Klebanov:1998hh}, is dual to a four-dimensional $\mc{N}=1$ SCFT arising from a stack of D3-branes sitting at the tip of a conifold. The Klebanov-Witten geometry is part of an infinite family of $AdS_5\times Y^{p,q}$ solutions \cite{Gauntlett:2004yd}, which were later shown to be dual to an infinite family of four-dimensional $\mc{N}=1$ quiver gauge theories \cite{Benvenuti:2004dy}. In the rest of this section, we will provide new, explicit ten-dimensional IIA uplifts of $D=5$ minimal gauged supergravity, by T-dualising the IIB uplifts on these manifolds along their $SU(2)$ directions.


\subsection{T-dualising the IIB uplift on $S^5$}

We will now apply the NATD technology discussed in Section \ref{section:NATD basics} to the Type IIB uplift of $D=5$ minimal gauged supergravity on $S^5$ (see Appendix \ref{sec:S5_truncation}). The matrix $M_{ij}$ \eqref{eq:Mij_matrix} and its inverse are given by
\begin{equation}
\begin{split}
M_{ij}&=\frac{1}{2g^2}\cos^2\xi\, \delta_{ij} + \epsilon_{ijk}x_k\,,\\
\left(M^{-1}\right)_{ij}&=\frac{2g^2\cos^2\xi\,\delta_{ij}+8g^6\cos^{-2}\xi\, x_{i}x_{j}-4 g^4\epsilon_{ijk}x_k}{\cos^4\xi+4g^4 r^2}\,,
\end{split}
\end{equation}
where we have set $v_i=x_i/\sqrt{2}$, and $r^2=x_1^2+x_2^2+x_3^2$. Using spherical polar coordinates,
\begin{equation}
x_1 = r\sin\theta\cos\phi\,,\quad x_2 = r\sin\theta\sin\phi\,,\quad x_3 = r\cos\theta\,,
\end{equation}
the T-dual metric becomes
\begin{align}\label{dualmetric_S5}
d\hat{s}^2_{10} &= ds^2_5+ \frac{1}{g^2}\left[d\xi^2 + \sin^2\xi\left(d\tau - \frac{2g}{3}A_{(1)} \right)^2\right]\nonumber \\
&\quad+ g^2\sec^2\xi\, dr^2+\frac{g^2r^2\cos^2\xi}{\cos^4\xi + 4g^4r^2}\left[d\theta^2+\sin^2\theta\left(d\phi-\frac{4g}{3}A_{(1)}\right)^2\right]\,,\numberthis
\end{align}
where $ds^2_5$ and $A_{(1)}$ are the metric and graviphoton of the five-dimensional theory. Next, the NSNS 2-form is
\begin{equation}\label{dualNSNS_S5}
\hat B_{(2)} =-\frac{2g^4r^3}{\cos^4\xi+4g^4r^2}\vol(S^2)+\frac{2g}{3}\left(\cos\theta dr-\frac{r\cos^4\xi\sin\theta}{\cos^4\xi + 4g^4r^2}d\theta\right) \wedge A_{(1)}\,,
\end{equation}
where $\vol(S^2) = \sin\theta\,d\theta\wedge d\phi$, and the dilaton is given by
\begin{equation}\label{dualdilaton_S5}
\hat{\Phi}=-\frac{1}{2}\log\left(\frac{4g^4r^2\cos^2\xi+\cos^6\xi}{8g^6}\right)\,.
\end{equation}
Turning now to the RR fluxes, we first write down 
\begin{equation}
\zeta_a=2g^2\delta_{ai}\sec^2\xi \,x_i\,,
\end{equation}
and
\begin{equation}
\Omega^{-1}=\frac{1}{\sqrt{1+4g^4r^2\sec^4\xi}}\left(\Gamma^{789}+2 g^2\sec^2\xi \,x_i\Gamma^i\right)\Gamma_{11}\,.
\end{equation}
From this, we find 
\begin{equation}\label{dualRR2_S5}
\hat F_{(2)} = -\frac{\sqrt{2}\cos^3\xi}{g^4}\sin\xi\,d\xi\wedge\left(d\tau - \frac{2g}{3}A_{(1)}\right) - \frac{\cos^4\xi}{3\sqrt{2}g^3}F_{(2)} \,,
\end{equation}
and
\begin{equation}\label{dualRR4_S5}
\begin{split}
\hat F_{(4)} &= -\frac{\cos^2\xi}{3\sqrt{2}g^2}\left[{\ast}_5F_{(2)}-\frac{1}{g}\sin^2\xi\,F_{(2)}\wedge \left(d\tau - \frac{2g}{3}A_{(1)} \right)\right]\wedge\left(\cos\theta\,dr-r\sin\theta\,d\theta\right) \\
&\quad +\frac{\sqrt{2}r\cos\xi\sin\xi\cos\theta}{3g^2}\left[{\ast}_5F_{(2)}- \frac{1}{g}F_{(2)}\wedge\left(d\tau - \frac{2g}{3}A_{(1)} \right)\right]\wedge d\xi\\
&\quad -\frac{2g^4r^3\sin\theta}{\cos^4\xi +4g^4r^2}\,\hat F_{(2)}\wedge d\theta\wedge\left(d\phi-\frac{4g}{3}A_{(1)}\right)\,.
\end{split}
\end{equation}
Here, $\hat F_{(2)}$ is the RR flux in the Type IIA theory, whereas $F_{(2)}$ is the flux of the five-dimensional theory. Also note that the application of the NATD procedure generates a six- and eight-form flux $\hat F_{(6)}$ and $\hat F_{(8)}$. These, however, as explained in the previous section, are dual to the two- and four-form fluxes presented above, $\hat F_{(6)} = -\hat{\ast}\hat F_{(4)}$, and $\hat F_{(8)} = \hat{\ast}\hat F_{(2)}$.

For a configuration where $A_{(1)} = 0$ and $ds^2_5$ is the metric on $AdS_5$, \eqref{dualmetric_S5}-\eqref{dualRR4_S5} recover the known non-Abelian T-dual of the $AdS_5\times S^5$ solution in \cite{Sfetsos:2010uq}.

\subsection{T-dualising the IIB uplift on $T^{1,1}$}

We will now apply the NATD technology to the Type IIB uplift on $T^{1,1}$ (see Appendix \ref{sec:T11_truncation}). The matrix $M_{ij}$ \eqref{eq:Mij_matrix} and its inverse are given by
\begin{equation}
\begin{split}
M_{ij}&= \frac{1}{3g^2}\delta_{ij} - \frac{1}{9g^2}\delta_{i3}\delta_{j3} + \epsilon_{ijk}x_k \,,\\
\left(M^{-1}\right)_{ij}&= \frac{6g^2\delta_{ij}+3g^2\delta_{i3}\delta_{j3}+81g^6x_ix_j-27g^4\epsilon_{ijk}x_k+9g^4\epsilon_{ij3}x_3}{2+9g^4\left[3\left(x_1^2+x_2^2\right)+2x_3^2\right]}\,,
\end{split}
\end{equation}
where we have set $v_i=x_i/\sqrt{2}$. Unlike the case with the $S^5$, the structure of $(M^{-1})_{ij}$ suggests the use of cylindrical polar coordinates
\begin{equation}
x_1 = \rho\cos\varphi \,,\quad x_2 = \rho\sin\varphi \,,\quad x_3 = x \,.
\end{equation}
In these coordinates, the T-dual metric becomes
\begin{equation}\label{dualmetric_T11}
\begin{split}
d\hat{s}^2_{10} &= ds^2_5 + \frac{1}{6g^2}d\Omega^2_2+ \frac{3g^2}{2\Pi}d\rho^2+ \frac{9g^2\Pi}{4\Pi+54g^4\rho^2}\left[dx + \frac{9g^4x\rho}{\Pi}d\rho\right]^2\\
&\quad + \frac{3g^2\rho^2}{2\Pi+27g^4\rho^2}\left[d\varphi+\cos\theta\,d\phi - 2gA_{(1)} \right]^2 \,,
\end{split}
\end{equation}
where we denote $\Pi = 1 + 9g^4x^2$, $d\Omega^2_2=d\theta^2+\sin^2\theta\,d\phi^2$, and for ease of notation, we set $\theta_1=\theta$ and $\phi_1=\phi$. Next, the NSNS 2-form is 
\begin{equation}\label{dualNSNS_T11}
\begin{split}
\hat B_{(2)} &= \frac{1}{(2\Pi+27g^4\rho^2)}\left(\cos\theta\,d\phi -2gA_{(1)}\right)\wedge\left(\Pi\,dx + 9g^4x\rho\,d\rho\right) \\
&\quad - \frac{9g^4}{2(2\Pi+27g^4\rho^2)}\left(2x\rho\,d\rho\wedge d\varphi + 3\rho^2d\varphi\wedge dx\right) \,,
\end{split}
\end{equation}
and the dilaton is given by
\begin{equation}\label{dualdilaton_T11}
\hat{\Phi}=-\frac{1}{2}\log\left(\frac{2\Pi+27g^4\rho^2}{81g^6}\right)\,.
\end{equation}
Turning now to the RR fluxes, we compute 
\begin{equation}
\zeta_a=3g^2\left(\frac{\sqrt{6}}{2}x_1,\frac{\sqrt{6}}{2}x_2,x_3\right)\,,
\end{equation}
and
\begin{equation}
\Omega^{-1}=\frac{\sqrt{2}}{\sqrt{2\Pi+27g^4\rho^2}}\left(\Gamma^{789} + \frac{3\sqrt{6}g^2}{2}x_1\Gamma^7+ \frac{3\sqrt{6}g^2}{2}x_2\Gamma^8+3g^2x_3\Gamma^9\right)\Gamma_{11}\,.
\end{equation}
From this, we find that
\begin{equation}\label{dualRR2_T11}
\hat F_{(2)} = \frac{2\sqrt{2}}{27g^4}\left(\vol(S^2) -gF_{(2)}\right) \,,
\end{equation}
where $\vol(S^2) = \sin\theta\,d\theta\wedge d\phi$, 
\begin{equation}\label{dualRR4_T11}
\begin{split}
\hat F_{(4)} &=-\frac{2x}{27g^3} F_{(2)}\wedge\vol(S^2)- \frac{\sqrt{2}}{9g^2}{\ast}_5F_{(2)}\wedge dx \\
&\quad-\frac{9g^4}{2(2\Pi + 27g^4\rho^2)}\hat F_{(2)}\wedge\left(2x\rho\,d\rho - 3\rho^2dx\right)\wedge\left(d\varphi+\cos\theta\,d\phi-2gA_{(1)}\right) \,,
\end{split}
\end{equation}
and we recall that the six and eight-form fluxes generated by the NATD procedure are dual to the two and four-form fluxes above, $\hat F_{(6)} = -\hat{\ast}\hat F_{(4)}$, and $\hat F_{(8)} = \hat{\ast}\hat F_{(2)}$.

For a configuration where $A_{(1)} = 0$ and $ds^2_5$ is the metric on $AdS_5$, \eqref{dualmetric_T11}-\eqref{dualRR4_T11} recover the known non-Abelian T-dual of the $AdS_5\times T^{1,1}$ solution in \cite{Itsios:2012zv,Itsios:2013wd}.

\subsection{T-dualising the IIB uplift on $Y^{p,q}$}

We will now turn the NATD crank one last time to the Type IIB uplift on $Y^{p,q}$ (see Appendix \ref{sec:Ypq_truncation}). The matrix $M_{ij}$ \eqref{eq:Mij_matrix} and its inverse are given by
\begin{equation}
M_{ij}= \frac{1-\tilde{y}}{3g^2}\delta_{ij} - \frac{1-\tilde{y}-6\Delta}{3g^2}\delta_{i3}\delta_{j3} + \epsilon_{ijk}x_k \,,
\end{equation}
and
\begin{equation}
\begin{split}
\left(M^{-1}\right)_{ij}&=\frac{g^2}{\Psi}\left[6\Delta\left(1-\tilde{y}\right)\delta_{ij}+\left[\left(1-\tilde{y}\right)^2-6\Delta\left(1-\tilde{y}\right)\right]\delta_{i3}\delta_{j3}\right.\\
&\left.\quad\,\,\quad+9g^4x_ix_j-3g^2\left(1-\tilde{y}\right)\epsilon_{ijk}x_k+3g^2\left(1-\tilde{y}-6\Delta\right)\epsilon_{ij3}x_3\right]\,,
\end{split}
\end{equation}
where we have set $v_i=x_i/\sqrt{2}$, and defined
\begin{equation}
\Psi=3g^4\left(1-\tilde{y}\right)\left(x_1^2+x_2^2\right)+2\Delta\left[\left(1-\tilde{y}\right)^2+9g^4x_3^2\right]\,.
\end{equation}
As with the case of $T^{1,1}$, the structure of $(M^{-1})_{ij}$ suggests the use of cylindrical polar coordinates
\begin{equation}
x_1 = \rho\cos\varphi \,,\quad x_2 = \rho\sin\varphi \,,\quad x_3 = x \,.
\end{equation}
In these coordinates, the T-dual metric becomes
\begin{equation}\label{dualmetric_Ypq}
\begin{split}
d\hat{s}^2_{10} &= ds^2_5+\frac{1}{g^2w(\tilde{y})v(\tilde{y})}d\tilde{y}^2+\frac{k(\tilde{y})^2}{g^2}\left[d\alpha-\frac{g}{3}A_{(1)}\right]^2\\
&\quad+ \frac{3g^2\left(1-\tilde{y}\right)}{2\Psi}\left(2\Delta+\frac{3g^4\rho^2\left(1-\tilde{y}\right)}{\left(1-\tilde{y}\right)^2+9g^4x^2}\right)d\rho^2\\
&\quad+\frac{3g^2\left(1-\tilde{y}\right)\rho^2}{\Psi}\left[\sqrt{\Delta}\,d\varphi+h\,d\alpha+\frac{2g(1-\tilde{y})}{9\sqrt{\Delta}}A_{(1)}\right]^2\\
&\quad+\frac{g^2\left[\left(1-\tilde{y}\right)^2+9g^4x^2\right]}{2\Psi}\left[dx+\frac{9g^4x\rho}{\left(1-\tilde{y}\right)^2+9g^4x^2}d\rho\right]^2\,,
\end{split}
\end{equation}
Next, the NSNS 2-form is 
\begin{equation}\label{dualNSNS_Ypq}
\begin{split}
\hat B_{(2)} &= \frac{\sqrt{\Delta}}{\Psi}\left(h\,d\alpha+\frac{2g(1-\tilde{y})}{9\sqrt{\Delta}}A_{(1)}\right)\wedge\left[\left((1-\tilde y)^2+9g^4x^2\right)dx + 9g^4x\rho\,d\rho\right]\\
&\quad - \frac{3g^4\rho}{2\Psi}\left(6x\Delta\,d\rho\wedge d\varphi + \left(1-\tilde{y}\right)\rho\,d\varphi\wedge dx\right) \,,
\end{split}
\end{equation}
and the dilaton is given by
\begin{equation}\label{dualdilaton_Ypq}
\hat{\Phi}=-\frac{1}{2}\log\left(\frac{\Psi}{9g^6}\right)\,.
\end{equation}
Turning now to the RR fluxes, we have
\begin{equation}
\zeta_a=3g^2\left(\frac{1}{\sqrt{6\Delta\left(1-\tilde{y}\right)}}x_1,\frac{1}{\sqrt{6\Delta\left(1-\tilde{y}\right)}}x_2,\frac{1}{1-\tilde{y}}x_3\right)\,,
\end{equation}
and
\begin{equation}
\begin{split}
\Omega^{-1}=&\frac{\sqrt{2\Delta}\left(1-\tilde{y}\right)}{\sqrt{\Psi}}\left(\Gamma^{789} + \frac{\sqrt{6}g^2}{2\sqrt{\Delta\left(1-\tilde{y}\right)}}x_1\Gamma^7\right.\\
&\left.\phantom{\frac{\sqrt{2\Delta}\left(1-\tilde{y}\right)}{\sqrt{\Psi}}( \,\,}+ \frac{\sqrt{6}g^2}{2\sqrt{\Delta\left(1-\tilde{y}\right)}}x_2\Gamma^8+\frac{3g^2}{1-\tilde{y}}x_3\Gamma^9\right)\Gamma_{11}\,.
\end{split}
\end{equation}
From this, we find
\begin{align*}\label{dualRR_Ypq}
\hat{F}_{(2)}&=-\frac{4(1-\tilde y)k}{3g^4}\sqrt{\frac{2\Delta}{wv}}d\tilde y\wedge\left(d\alpha-\frac{g}{3}A_{(1)}\right) - \frac{2\sqrt{2}(1-\tilde y)}{3g^2}\left(\Delta-\frac{wf}{6}\right)F_{(2)}
\,,\\
\hat{F}_{(4)}&= -\frac{2\sqrt{2\Delta}(1-\tilde y)}{9g^3}{\ast}_5F_{(2)}\wedge\hat e^{9}+ \frac{\sqrt{2wv}(1-\tilde y)}{27g^3}F_{(2)}\wedge\hat{e}^{69}  \\
&\quad  -\frac{2}{3g}\sqrt{\frac{1-\tilde y}{3}}\left(x_1{\ast}_5F_{(2)}\wedge\hat{e}^{8} - x_2{\ast}_5F_{(2)}\wedge\hat{e}^{7} + \sqrt{\frac{6\Delta}{1-\tilde y}}\,x_3F_{(2)}\wedge\hat{e}^{56}\right) \numberthis \\
&\quad +\frac{1}{27g}\sqrt{\frac{3wv(1-\tilde y)}{\Delta}}\left(x_1F_{(2)}\wedge\hat{e}^{68}-x_2F_{(2)}\wedge\hat{e}^{67} -  \sqrt{\frac{6\Delta}{1-\tilde y}}\,x_3{\ast}_5F_{(2)}\wedge\hat{e}^{5}\right)\\
&\quad-g^2\sqrt{\frac{3}{2\Delta(1-\tilde y)}}\hat F_{(2)}\wedge\left(x_1\hat e^{89}-x_2\hat e^{79} + \sqrt{\frac{6\Delta}{1-\tilde y}}\,x_3\hat e^{78}\right)
\,.
\end{align*}
Here, $\hat{e}^{5,6} = e^{5,6}$ in \eqref{eq:pure_original_vielbeins_Ypq}, and $\hat e^{7,8,9}$ are the right-moving frames $\hat e_+$ in \eqref{leftrightvielbeins}. Again, the six and eight-form fluxes generated are dual to the two- and four-form fluxes, $\hat F_{(6)} = -\hat{\ast}\hat F_{(4)}$, and $\hat F_{(8)} = \hat{\ast}\hat F_{(2)}$.

For a configuration where $A_{(1)} = 0$ and $ds^2_5$ is the metric on $AdS_5$, \eqref{dualmetric_Ypq}-\eqref{dualRR_Ypq} recover the known non-Abelian T-dual of the $AdS_5\times Y^{p,q}$ solution in \cite{Sfetsos:2014tza}.

\subsection{Some comments}


\noindent All of the fluxes of the Type IIA theory are turned on in all three embeddings. We thus expect that the original D3-brane is replaced by a configuration of D4-, NS5-, and D6-branes. Unlike the Abelian case, the charges of the branes in the dual configuration are not straightforwardly defined. This is because the dualised internal space $M_5$ is \textit{a priori} non-compact. For the $S^5$ case, the non-compact coordinate is the radius $r$, and for the $T^{1,1}$ and $Y^{p,q}$ cases, the non-compact coordinate is $\rho$. Due to this non-compactness, the charge of the NS5-brane, given by the integral of $\hat H_{(3)} = d\hat B_{(2)}$, which always has a direction in either $r$ or $\rho$, is not well-defined if we want to embed this system in a string setting.\footnote{A purely classical, supergravity system does not impose quantisation requirements on the integrals of the fluxes.} A remedy for this problem was proposed in \cite{Sfetsos:2014tza,Lozano:2013oma}, where the range of the apparent non-compact coordinates $r$ and $\rho$ can be fixed by requiring
\begin{equation}\label{eq:some_b0_term}
\hat b_0 = \frac{1}{4\pi^2}\int_{\Sigma_2}\hat B_{(2)} \in [0,1]\,,
\end{equation}
where $\Sigma_2$ is a non-trivial two-cycle. For the $SU(2)$ T-dual background of $AdS_5\times S^5$, it was shown in \cite{Lozano:2016kum} that \eqref{eq:some_b0_term} corresponds to the charge of the NS5-brane for $r\in[0,n\pi]$ with $n\in \mathbb{N}$. For the $T^{1,1}$ and $Y^{p,q}$ cases, there is an additional coordinate $x$ corresponding to the $x_3$-axis of the cylindrical polar system that appears non-compact, but the integration of the volume form over $x$ does not give an infinite contribution to the volume of the internal space, so it is not a genuine non-compact direction. 

Even though the dualised configuration in IIA contains a much richer brane content than the original IIB setup, we can show that the dualisation procedure considered here does not break any further supersymmetries. From \cite{Kelekci:2014ima}, the task of determining the amount of preserved supersymmetry of a T-dualised (Abelian or non-Abelian) solution is boiled down to computing the action of the Kosmann derivative of the Killing spinor in the original, non-dualised solution, along the dualised directions. If the Kosmann derivative vanishes, the T-dualised solution will not break any further supersymmetries. Explicitly, the action of the Kosmann derivative along a Killing vector $K$ on a spinor $\epsilon$ is given by\cite{Kosmann:1972,Ortin:2002qb}
\begin{equation}
\mc{L}_K\epsilon = \nabla_K\epsilon + \frac{1}{8}(dK)_{MN}\Gamma^{MN}\epsilon \,.
\end{equation}
For our purposes, we would need to compute the Kosmann derivative of the Type IIB Killing spinor $\epsilon$ with respect to the right-invariant vector fields of $SU(2)$,
\begin{align}
\begin{split}\label{eq:SU(2)_right_vec_fields}
K_1&=-\frac{\cos\phi}{\sin\theta}\,\partial_\psi+\sin\phi\,\partial_\theta+\cot\theta\cos\phi\,\partial_\phi\,,\\
K_2&=\frac{\sin\phi}{\sin\theta}\,\partial_\psi+\cos\phi\,\partial_\theta-\cot\theta\sin\phi\,\partial_\phi\,,\\
K_3&=\partial_\phi\,.
\end{split}
\end{align}
The expressions for the Killing spinor for the $S^5$, $T^{1,1}$ and $Y^{p,q}$ cases are given in Appendix \ref{sec:IIB_truncations_on_SE5}. It is a straightforward, albeit tedious calculation, to show that the Kosmann derivative indeed vanishes in all our cases. 
This result should not come as a surprise given that the IIB Killing spinors are independent of the $SU(2)$ directions $(\psi,\theta,\phi)$ (see Appendix \ref{sec:IIB_truncations_on_SE5}). As discussed in \cite{Kelekci:2014ima}, if the Killing spinor $\epsilon$ is constant with respect to the $SU(2)$ directions, the Kosmann derivative is guaranteed to vanish.\footnote{Note that this is a frame dependent statement, as spinors transform linearly under a local Lorentz transformation. A more precise statement is that the Kosmann derivative vanishes if there is a frame in which the Killing spinor is constant along the $SU(2)$ directions.} Thus we can conclude that the dual IIA uplifts do not break any further supersymmetries.

We also note that the chirality of the theory flips/remain the same when the dimension of the isometry group is odd/even under NATD \cite{Sfetsos:2010uq,Kelekci:2014ima}.\footnote{This result is a generalisation of that in  \cite{Hassan:1999bv}, where under Abelian T-duality, the chirality of the theory undergoes a flip.} Schematically, the Majorana-Weyl Killing spinor in IIA can be written as
\begin{equation}
\hat\epsilon = \begin{pmatrix}\hat\epsilon_1 \\ \hat\epsilon_2 \end{pmatrix} \,,
\end{equation}
and after T-duality, the spinor will undergo a transformation \cite{Kelekci:2014ima},
\begin{equation}\label{eq:spinor_transformation}
\epsilon_1=\Omega^{-1}\hat{\epsilon}_1\,,\quad \epsilon_2=\hat{\epsilon}_2\,.
\end{equation}
In our case where the isometry group is $SU(2)$, the overall chirality of the ten-dimensional theory flips after T-duality. 


Finally, we want to comment on the curvature singularity in the $S^5$ case. Similar to the $SU(2)$ T-dual of $AdS_5\times S^5$ in \cite{Sfetsos:2010uq}, our T-dual metric \eqref{dualmetric_S5} is plagued by a curvature singularity at $\xi = \tfrac{\pi}{2}$, whereas the $SU(2)$ T-dual backgrounds for $N_5=T^{1,1},Y^{p,q}$ are actually free of such curvature singularities. This begs the question: why is it only the $S^5$ case that has a singularity? The metric on a round $S^5$ has an $SO(6)$ isometry. In order to apply the NATD transformations, we have to parameterise the $S^5$ in a way such that an $SU(2)$ isometry becomes manifest. This is done in \eqref{S5_parameters} by choosing the $(\xi,\tau,\psi,\theta,\phi)$ coordinates on $S^5$. Here, $(\xi,\tau)$ parameterise an $S^2/\mb{Z}_2\equiv\mb{R}\text{P}^2$ with isometry $SO(3)/\mb{Z}_2$,\footnote{The $\mb{Z}_2$ identification is because the range of $\xi$ only covers the Northern hemisphere of the $S^2$.} and $(\psi,\theta,\phi)$ are the Euler angles parameterising an $S^3$, which has an $SU(2)^2$ isometry. In these coordinates, however, there are coordinate singularities at $\xi=0$ and $\xi=\tfrac{\pi}{2}$. The $\xi=0$ coordinate singularity is a bolt singularity, and is removable provided $\tau\in[0,2\pi)$, but the $\xi=\tfrac{\pi}{2}$ coordinate singularity is only removable by coordinate transformations that mixes the $\mb{R}\text{P}^2$ and $S^3$ coordinates in a way that essentially re-introduces the whole $SO(6)$ isometry. The existence of these coordinate singularities can also be interpreted in terms of the isometry groups by the fact that $SO(3)/\mb{Z}_2\times SU(2)^2$ is only a subgroup of $SO(6)$, and since $SO(6)$ is simple, it can never be written as a product of these subgroups. Once we have dualised one of the $SU(2)$ factors, we have lost the mixed coordinate transformations that resolves the $\xi=\tfrac{\pi}{2}$ coordinate singularity, which promotes it to an actual curvature singularity. For the $T^{1,1}$ and $Y^{p,q}$ cases, their isometry groups, up to discrete identifications, naturally factorise into $SU(2)\times U(1)^2$, so any apparent coordinate singularities are cured by coordinate transformations that only mix within the $SU(2)$ and $U(1)^2$ factors separately. As such, the resulting metric after T-dualising one of the $SU(2)$ directions, is still smooth.

\section{T-dual spindle solutions in IIA}\label{section:NATD_S5}

In the previous section, we have constructed new, explicit Type IIA uplifts of $D=5$ minimal gauged supergravity via NATD. Clearly, these formulae can be applied to construct Type IIA solutions by uplifting solutions of $D=5$ minimal gauged supergravity. There are numerous known supersymmetric solutions of $D=5$ minimal gauged supergravity, notably such as the maximally supersymmetric $AdS_5$ geometry and the toplogically twisted $AdS_3\times \Sigma_{\mathfrak{g}}$ geometry \cite{Klemm:2000nj} (i.e. $\Sigma_{\mathfrak{g}}$ is a Riemann surface with $\mathfrak{g}>1$ obtained by taking a discrete quotient of $\mathbb{H}^2$), which can all be fitted into the classification results in \cite{Gauntlett:2003fk}.

In this section, we will not attempt to cover and uplift all these results. Instead, we will apply our Type IIA uplifts to the recently discovered $AdS_3\times \Sigma$ solution in \cite{Ferrero:2020laf}, which give rise to Type IIA solutions of form $AdS_3\times \Sigma \times M_5$ respectively. In the following, we will first provide a review of the supersymmetric $AdS_3\times \Sigma$ solution in \cite{Ferrero:2020laf}, then we will proceed to provide the explicit expressions of these $AdS_3\times \Sigma \times M_5$ solutions, and finally we conclude this section with some remarks about the solutions.

\subsection{Review of the supersymmetric $AdS_3\times \Sigma$ solution}

The supersymmetric $AdS_3\times \Sigma$ solution in \cite{Ferrero:2020laf} describing the near-horizon limit of a D3-brane wrapping on the spindle $\Sigma$ is given by\footnote{The coupling $g$ is set to 1 in \cite{Ferrero:2020laf}.}
\begin{equation}\label{D3spindle}
ds^2_5 = \frac{4y}{9g^2}ds^2_{AdS_3} + \frac{1}{g^2}ds^2_{\Sigma} \,,\quad {A}_{(1)} = \frac{1}{4g}\left(1-\frac{a}{y}\right)dz \,,
\end{equation}
where the metric on $AdS_3$ has unit radius, and
\begin{equation}\label{spindlemetric}
ds^2_{\Sigma} = \frac{y}{q(y)}dy^2 + \frac{q(y)}{36y^2}dz^2 \,,\quad q(y) = 4y^3-9y^2+6ay-a^2\,,\quad a\in(0,1)\,.
\end{equation}
Since $a\in(0,1)$, the three roots $y_i$ of $q(y)$ are real and positive. By labelling the roots $y_1<y_2<y_3$, \eqref{spindlemetric} is a positive definite metric on $\Sigma$ provided $y\in[y_1,y_2]$. Near $y_1$ and $y_2$, it is impossible to choose a single period for the $U(1)$ coordinate $z$ to remove the conical singularities. As shown in \cite{Ferrero:2020laf}, if
\begin{equation}
a = \frac{(n_--n_+)^2(2n_-+n_+)^2(n_-+2n_+)^2}{4(n_-^2+n_-n_++n_+^2)^3} \,,\quad z \sim z + \frac{4\pi(n_-^2+n_-n_++n_+^2)}{3n_-n_+(n_-+n_+)} \,,
\end{equation}
where $n_->n_+$ are coprime, then the metric \eqref{spindlemetric} describes a spindle $\Sigma=\mathbb{WCP}^1_{[n_-,n_+]}$ with conical deficit angles $2\pi(1-1/n_\mp)$ at $y_1$ and $y_2$ respectively. The expressions for the flux $F_{(2)} =dA_{(1)}$ and its Hodge dual are required before applying the uplift formulae we derived in the previous section, so for later convenience, we will give them here
\begin{equation}
F_{(2)} = \frac{3a}{2gy^{3/2}}\vol(\Sigma) \,,\quad {\ast}_5F_{(2)} = \frac{4a}{9g^2}\vol(AdS_3) \,.
\end{equation}
From \cite{Ferrero:2020laf}, the Killing spinor of the spindle solution \eqref{D3spindle} is given by
\begin{equation}
\chi=\vartheta_{AdS_3}\otimes \vartheta_{\Sigma}\,,
\end{equation}
where $\vartheta_{AdS_3}$ is the Killing spinor on $AdS_3$, and
\begin{equation}\label{eq:spindle_spinor}
\vartheta_{\Sigma}=\left(\frac{\sqrt{q_1(y)}}{\sqrt{y}},i\frac{\sqrt{q_2(y)}}{\sqrt{y}}\right)\,,
\end{equation}
where
\begin{equation}
q_1(y)=-a+2y^{3/2}+3y\,,\quad q_2(y)=a+2y^{3/2}-3y 
\end{equation}
satisfy $q(y)=q_1(y)q_2(y)$. It is important to emphasise again that the supersymmetry of this solution is not realised with the usual topological twist, as seen from the fact that $\theta_\Sigma$ is not covariantly constant on $\Sigma$.

In the rest of this section, we will present the T-dual solutions in Type IIA using the three embeddings of $D=5$ minimal gauged supergravity discussed in Section \ref{section:IIAembedding}. Since the solutions are obtained by a straightforward substitution of the $ds^2_5$ and $A_{(1)}$ in \eqref{D3spindle}, we will only present the NSNS sector, noting that the RR fluxes can be obtained easily using the formulae in Section \ref{section:IIAembedding}. From the general result in Section \ref{section:IIAembedding}, these T-dual solutions are supersymmetric, and are related to some two-dimensional $\mathcal{N}=(0,2)$ SCFTs from a holographic perspective.

\subsection{NATD of $AdS_3\times \Sigma\times N_5$}

\subsubsection{Case 1: $S^5$}

\noindent Using \eqref{dualmetric_S5}, the T-dual metric is 
\begin{align}\label{spindlemetric_S5}
d\hat{s}^2_{10} &= \frac{4y}{9g^2}ds^2_{AdS_3} + \frac{1}{g^2}\left[\frac{y}{q(y)}dy^2 + \frac{q(y)}{36y^2}dz^2\right] + \frac{1}{g^2}\left[d\xi^2 + \sin^2\xi\left(d\tau - \frac{1}{6}\left(1-\frac{a}{y}\right)dz \right)^2\right]\nonumber \\
&\quad+ g^2\sec^2\xi\, dr^2+\frac{g^2r^2\cos^2\xi}{\cos^4\xi + 4g^4r^2}\left[d\theta^2+\sin^2\theta\left(d\phi-\frac{1}{3}\left(1-\frac{a}{y}\right)dz\right)^2\right]\,.\numberthis
\end{align}
Next, from \eqref{dualNSNS_S5}, the NSNS 2-form is
\begin{equation}\label{spindleNSNS_S5}
\hat B_{(2)} =-\frac{2g^4r^3}{\cos^4\xi+4g^4r^2}\vol(S^2)+\frac{1}{6}\left(1-\frac{a}{y}\right)\left(\cos\theta dr-\frac{r\cos^4\xi\sin\theta}{\cos^4\xi + 4g^4r^2}d\theta\right) \wedge dz\,,
\end{equation}
where $\vol(S^2) = \sin\theta\,d\theta\wedge d\phi$, and the dilaton is given by \eqref{dualdilaton_S5}
\begin{equation}\label{spindledilaton_S5}
\hat{\Phi}=-\frac{1}{2}\log\left(\frac{4g^4r^2\cos^2\xi+\cos^6\xi}{8g^6}\right)\,.
\end{equation}
The RR fluxes are given by \eqref{dualRR2_S5} and \eqref{dualRR4_S5}. 

\subsubsection{Case 2: $T^{1,1}$}

\noindent Using \eqref{dualmetric_T11}, the T-dual metric is
\begin{equation}\label{spindlemetric_T11}
\begin{split}
d\hat{s}^2_{10} &= \frac{4y}{9g^2}ds^2_{AdS_3} + \frac{1}{g^2}\left[\frac{y}{q(y)}dy^2 + \frac{q(y)}{36y^2}dz^2\right] + \frac{1}{6g^2}\left[d\theta^2+\sin^2\theta\,d\phi^2\right]  \\
&\quad+ \frac{3g^2}{2\Pi}d\rho^2 + \frac{3g^2\rho^2}{2\Pi+27g^4\rho^2}\left[d\varphi+\cos\theta\,d\phi - \frac{1}{2}\left(1-\frac{a}{y}\right)dz \right]^2  \\
&\quad + \frac{9g^2\Pi}{4\Pi+54g^4\rho^2}\left[dx + \frac{9g^4x\rho}{\Pi}d\rho\right]^2 \,,
\end{split}
\end{equation}
where we recall $\Pi = 1 + 9g^4x^2$. Next, the NSNS 2-form from \eqref{dualNSNS_T11} is 
\begin{equation}\label{spindleNSNS_T11}
\begin{split}
\hat B_{(2)} &= \frac{1}{(2\Pi+27g^4\rho^2)}\left(\cos\theta\,d\phi - \frac{1}{2}\left(1-\frac{a}{y}\right)dz\right)\wedge\left(\Pi\,dx + 9g^4x\rho\,d\rho\right) \\
&\quad - \frac{9g^4}{2(2\Pi+27g^4\rho^2)}\left(2x\rho\,d\rho\wedge d\varphi + 3\rho^2d\varphi\wedge dx\right) \,,
\end{split}
\end{equation}
and the dilaton \eqref{dualdilaton_T11} is given by
\begin{equation}\label{spindledilaton_T11}
\hat{\Phi}=-\frac{1}{2}\log\left(\frac{2\Pi+27g^4\rho^2}{81g^6}\right)\,.
\end{equation}
The RR fluxes are given in \eqref{dualRR2_T11} and \eqref{dualRR4_T11}. 

\subsubsection{Case 3: $Y^{p,q}$}

\noindent Using \eqref{dualmetric_Ypq}, the T-dual metric is
\begin{align*}\label{spindlemetric_Ypq}
&d\hat{s}^2_{10} = \, \frac{4y}{9g^2}ds^2_{AdS_3} + \frac{1}{g^2}\left[\frac{y}{q(y)}dy^2 + \frac{q(y)}{36y^2}dz^2\right]+\frac{1}{g^2w(\tilde{y})v(\tilde{y})}d\tilde{y}^2+\frac{k(\tilde{y})^2}{g^2}\left(d\alpha-\frac{1}{12}\left(1-\frac{a}{y}\right)dz\right)^2\\
&+\frac{g^2\left[\left(1-\tilde{y}\right)^2+9g^4x^2\right]}{2\Psi}\left(dx+\frac{9g^4x\rho}{\left(1-\tilde{y}\right)^2+9g^4x^2}d\rho\right)^2+ \frac{3g^2\left(1-\tilde{y}\right)}{2\Psi}\left(2\Delta+\frac{3g^4\rho^2\left(1-\tilde{y}\right)}{\left(1-\tilde{y}\right)^2+9g^4x^2}\right)d\rho^2\\
&+\frac{3g^2\left(1-\tilde{y}\right)\rho^2}{\Psi}\left(\sqrt{\Delta}\,d\varphi+h\,d\alpha+\frac{1-\tilde{y}}{18\sqrt{\Delta}}\left(1-\frac{a}{y}\right)dz\right)^2\,,\numberthis
\end{align*}
where we recall
\begin{equation}
\Psi=3g^4\left(1-\tilde{y}\right)\rho^2+2\Delta\left[\left(1-\tilde{y}\right)^2+9g^4x^2\right]\,.
\end{equation}
Next, from \eqref{dualNSNS_Ypq}, the NSNS 2-form is
\begin{equation}\label{spindleNSNS_Ypq}
\begin{split}
\hat B_{(2)} &= \frac{\sqrt{\Delta}}{\Psi}\left(h\,d\alpha+\frac{1-\tilde{y}}{18\sqrt{\Delta}}\left(1-\frac{a}{y}\right)dz\right)\wedge\left[\left((1-\tilde y)^2+9g^4x^2\right)dx + 9g^4x\rho\,d\rho\right]\\
&\quad - \frac{3g^4\rho}{2\Psi}\left(6x\Delta\,d\rho\wedge d\varphi + \left(1-\tilde{y}\right)\rho\,d\varphi\wedge dx\right) \,,
\end{split}
\end{equation}
and the dilaton \eqref{dualdilaton_Ypq} is given by
\begin{equation}\label{spindledilaton_Ypq}
\hat{\Phi}=-\frac{1}{2}\log\left(\frac{\Psi}{9g^6}\right)\,.
\end{equation}
The RR fluxes are given in \eqref{dualRR_Ypq}.

\subsection{Some comments}\label{se:commentsS5}

Although we have not constructed the Type IIA Killing spinors directly, we can, using the arguments presented in the previous section, deduce that the dual field theories indeed exhibit $d=2$, $\mathcal{N}=(0,2)$ supersymmetry. From Appendix \ref{sec:IIB_truncations_on_SE5}, we find that the Type IIB Killing spinor of the spindle solution has the form
\begin{equation}
\epsilon = \theta_{AdS_3}\otimes\theta_\Sigma\otimes\varepsilon\,,
\end{equation}
where $\varepsilon$ is an $SU(2)$ invariant spinor. Now, the transformation matrix $\Omega^{-1}$ \eqref{eq:omega_matrix_we_use}, which transforms the IIB Killing spinor into the IIA Killing spinor, only acts on the T-dualised directions which are transverse to the $D=5$ spacetime, so it only touches $\varepsilon$. As a result, the dual IIA Killing spinor can be repackaged into the form
\begin{equation}
\hat\epsilon = \theta_{AdS_3}\otimes\theta_\Sigma\otimes\hat\varepsilon\,.
\end{equation}
The corresponding spindle solution in IIA is then holographically related to an $d=2$ SCFT with $\mathcal{N}=(0,2)$ supersymmetry.

Now we would like to make some specific comments on the holographic aspects of our Type IIA solutions. The spindle solution \eqref{D3spindle} uplifted to Type \Romannum{2}B on $S^5$ describes the near-horizon limit of D3-branes wrapping a spindle $\Sigma$. When the parameter $a$ in \eqref{D3spindle} is set to $0$, the standard maximally supersymmetric $AdS_5\times S^5$ solution, dual to the $\mc{N}=4$ SYM theory, is recovered. The $SU(2)$ T-dual of $AdS_5\times S^5$ was first discovered and discussed in \cite{Sfetsos:2010uq}, and the M-theory uplift of this T-dual solution falls into the class of $\mc{N}=2$ Gaiotto-Maldacena backgrounds \cite{Gaiotto:2009gz,Reid-Edwards:2010vpm} dual to $d=4$, $\mc{N}=2$ SCFTs discussed in \cite{Gaiotto:2009we}.\footnote{However, the solution in \cite{Sfetsos:2010uq} has a curvature singularity, which we have discussed in Section \ref{section:IIAembedding}. This makes an exact identification difficult at the field theory level.} For $a\neq 0$, the IIB spindle solution uplifted on $S^5$ is proposed as the IR limit of the $\mc{N}=4$ SYM compactified on a spindle \cite{Ferrero:2020laf}. Analogously, this should suggest that our T-dual IIA solution \eqref{spindlemetric_S5}-\eqref{spindledilaton_S5} is holographically related to a two-dimensional $\mathcal{N}=(0,2)$ SCFT and should capture features of a $d=4$, $\mc{N}=2$ SCFT compactified on a spindle. Meanwhile, in both the $T^{1,1}$ and $Y^{p,q}$ cases, the T-dualised $AdS_5$ solutions associated with these manifolds are suggested to be connected to the $d=4$, $\mc{N}=1$ theories considered in \cite{Bah:2012dg}. More specifically for the $T^{1,1}$ case, it was shown explicitly in \cite{Itsios:2013wd} that the M-theory uplift of the T-dualised solution fits within the ansatz considered in \cite{Bah:2012dg}, and hence is associated with the description of M5-branes wrapping over a two-sphere. Based on these known results, we suggest that our T-dual solutions in \eqref{spindlemetric_T11}-\eqref{spindledilaton_T11} and \eqref{spindlemetric_Ypq}-\eqref{spindledilaton_Ypq} are holographically dual to two-dimensional $\mathcal{N}=(0,2)$ SCFTs, arising as the IR limits of $d=4$, $\mc{N}=1$ theories compactified on a spindle.

\section{Final comments}\label{conclusion}

In this note, we have constructed new, explicit ten-dimensional IIA uplifts of $D=5$ minimal gauged supergravity, by T-dualising the known IIB uplifts on $S^5$, $T^{1,1}$ and $Y^{p,q}$ respectively, along their $SU(2)$ isometries. The internal geometries of the IIA uplifts are \textit{a priori} non-compact, but can be compactified by requiring consistency of the string theory embedding following the methods of \cite{Sfetsos:2014tza,Lozano:2013oma}. In the $S^5$ case, we find that the internal geometry also contains a curvature singularity which was first noted in the $SU(2)$ T-dual solution of $AdS_5\times S^5$ in \cite{Sfetsos:2010uq}. The nature of the singularity and the reason why it only appears in the $S^5$ but not the $T^{1,1}$ and $Y^{p,q}$ cases is discussed. 

One can apply our formulae to construct new Type IIA solutions by uplifting existing solutions of $D=5$ minimal gauged supergravity. In this note, we have applied these methods to construct new, explicit Type IIA solutions of form $AdS_3\times \Sigma \times M_5$, where $\Sigma$ is the weighted projective space $\mathbb{WCP}^1_{[n_-,n_+]}$, also known as a spindle, by uplifting the $AdS_3\times\Sigma$ solution of \cite{Ferrero:2020laf}. The T-dual solutions do not break any further symmetries, and we suggest that these solutions are holographically related to $d=2$, $\mathcal{N}=(0,2)$ SCFTs which should arise from compactifying certain $d=4$ SCFTs on a spindle. It would be extremely interesting to work out the exact field theory mechanism and confirm our proposal, which for now, we will leave as an intriguing open question.

One direct extension of this work is to apply NATD to the Type IIB uplift of the $U(1)^3$ gauged supergravity in five dimensions on $S^5$, which can be obtained by relaxing the conditions in Appendix \ref{sec:S5_truncation} to allow three distinct Cartan gauge fields ($A^{12}_{(1)}, A^{34}_{(1)}, A^{56}_{(1)}$) and two extra scalar fields (namely $X_1$ and $X_2$). This $D=5$ $U(1)^3$ theory provides a more general description of wrapped D3-branes. A particular solution of the model that is of interest is the multi-charge spindle solution presented in \cite{Boido:2021szx}, which is interpreted as the near-horizon limit of D3-branes wrapping the spindle, but with extra parameters determining the twisting of the $S^5$ over the spindle when compared to the single charge spindle solution. It is holographically dual to a two-dimensional $\mathcal{N}=(0,2)$ SCFT, arising as the IR limit of the $\mc{N}=4$ SYM compactified on a spindle. The dual description can be enhanced to a $\mathcal{N}=(2,2)$ SCFT in a special case, which describes a configuration of D3-branes wrapped on a topological disk. More details on this are given in \cite{Couzens:2021tnv,Suh:2021ifj}. In terms of the amount of supersymmetry preserved after performing the NATD, we expect the same argument in Section \ref{section:IIAembedding} to hold for the multi-charge spindle solution, so that no further supersymmetry is broken. The $SU(2)$ T-dual of the IIB embedding of the $D=5$ $U(1)^3$ model, albeit being a rather involved algebraic exercise, would give rise to a new realisation of the model in Type IIA, and consequently, new supersymmetric spindle solutions with potentially interesting holographic properties.

In comparison with the $S^5$ and $Y^{p,q}$ cases, we notice that the explicit T-dualised uplift on $T^{1,1}$ is remarkably clean and free of any curvature singularity, which of course have to do with the simplicity of the $T^{1,1}$ metric and its immediate $SU(2)$-manifest topology. The holographic interpretation for the $T^{1,1}$ case is also clearer since the gravity solution can be smoothly fitted into the framework of \cite{Bah:2012dg}. These observations suggest another extension of our work, that is by applying the NATD procedure to the truncation in \cite{Cassani:2010na,Bena:2010pr}, which is a $D=5$, $\mathcal{N}=4$ supergravity coupled to three vector multiplets and obtained by reducing Type IIB supergravity on $T^{1,1}$. By carrying out the NATD procedure along the $SU(2)$-directions, one would obtain new embedding of this five-dimensional theory in massive Type IIA supergravity,\footnote{All of the Type IIB NSNS and RR fluxes are switched on in the consistent truncation on $T^{1,1}$ \cite{Cassani:2010na,Bena:2010pr}, hence the T-dualised theory is expected to live within massive Type IIA supergravity.} which can be of great interest from the perspectives of holography and supergravity.

\subsection*{Acknowledgments}

We thank Jerome Gauntlett and Christopher Rosen for discussions. KCMC is supported by an Imperial College President's PhD Scholarship. 

\begin{appendix}

\section{Conventions for IIA and IIB}

In this section, we provide the Bianchi identities and equations of motion of both Type IIA and Type IIB supergravities which we use throughout this note.

\subsection{Type IIA}

The field content of Type IIA supergravity contains a metric $\hat g$, which we will take to be the string frame metric, a dilaton $\hat\Phi$, the NSNS 2-form $\hat B_{(2)}$, and the RR fluxes $\hat F_{(2)}$ and $\hat F_{(4)}$. The Bianchi identities are given by
\begin{equation}
d\hat F_{(2)} = 0 \,,\quad d\hat H_{(3)} = 0 \,,\quad d\hat F_{(4)}  - \hat H_{(3)}\wedge\hat F_{(2)} = 0 \,,
\end{equation}
with $\hat H_{(3)} = d\hat B_{(2)}$. The flux equations of motion are
\begin{equation}
\begin{split}
&d{\hat\ast}\hat F_{(2)} + \hat H_{(3)}\wedge{\hat\ast}\hat F_{(4)} = 0 \,, \\
&d\left( e^{-2\hat\Phi}{\hat\ast}\hat H_{(3)}\right) - \hat F_{(2)}\wedge{\hat\ast}\hat F_{(4)} - \frac{1}{2}\hat F_{(4)}\wedge \hat F_{(4)} = 0\,,\\
&d{\hat\ast}\hat F_{(4)} + \hat H_{(3)}\wedge\hat F_{(4)} = 0 \,.
\end{split}
\end{equation}
The dilaton equation is 
\begin{equation}
\hat\Box\hat\Phi-(\partial\hat\Phi)^2+\frac{1}{4}\hat R -\frac{1}{48}(\hat H_{(3)})^2 = 0 \,,
\end{equation}
and the Einstein equations are
\begin{equation}
\begin{split}
\hat R_{MN} &= -2\hat\nabla_M\hat\nabla_N\hat\Phi + \frac{1}{4}\hat H^{}_{MPQ}\hat H_N^{\ph{N}PQ} +\frac{e^{2\hat\Phi}}{2}\left(\hat F^{}_{MP}\hat F_N^{\ph{N}P}-\frac{1}{4}\hat g_{MN}(\hat F_{(2)})^2\right) \\
&\quad+\frac{e^{2\hat\Phi}}{12}\left(\hat F^{}_{MPQR}\hat F_N^{\ph{N}PQR} - \frac{1}{8}\hat g_{MN}(\hat F_{(4)})^2\right) \,.
\end{split}
\end{equation}

\subsection{Type IIB}

In this note, we work with a consistent truncation of Type IIB supergravity to just the metric and the 5-form flux (i.e. $\Phi=C_{(0)}=B_{(2)}=C_{(2)}=0$). The 5-form flux $F_{(5)}$ is self-dual 
\begin{equation}
F_{(5)} = {\ast}F_{(5)}  \,,
\end{equation}
and satisfies the Bianchi identity
\begin{equation}
dF_{(5)} =  0 \,.
\end{equation}
The Einstein equations are
\begin{equation}
R_{MN} = \frac{1}{96}F_{MPQRS}^{}F_{N}^{\ph{N}PQRS} \,.
\end{equation}

\section{Truncations of IIB supergravity to $D=5$ minimal gauged supergravity}\label{sec:IIB_truncations_on_SE5}

In this section, we will summarise the consistent truncations of Type IIB supergravity on $S^5$, $T^{1,1}$ and $Y^{p,q}$ respectively to the five-dimensional minimal gauged supergravity. The $S^5$ truncation is a well-known example \cite{Cvetic:2000nc}, however, we notice that some of the key details of this truncation are often only briefly discussed in the literature. Hence we will discuss the $S^5$ truncation in detail in Appendix \ref{sec:S5_truncation}. For both the $T^{1,1}$ and $Y^{p,q}$ truncations, we will summarise the key ingredients of the truncations presented in \cite{Buchel:2006gb} in Appendices \ref{sec:T11_truncation}-\ref{sec:Ypq_truncation}, as well as provide new derivations of the Killing spinors in a frame where the $SU(2)$ isometries of $T^{1,1}$ and $Y^{p,q}$ are manifest. In this frame, we find that the Killing spinors are, unsurprisingly, invariant under the $SU(2)$ isometries, which is particularly useful in analysing the preserved supersymmetry of the T-dual solutions.

\subsection{$S^5$ truncation}\label{sec:S5_truncation}
Using the results of \cite{Cvetic:2000nc}, the $S^5$ truncation of type \Romannum{2}B supergravity to $D=5$ minimal gauged supergravity, at the level of the bosonic fields, is given by
\begin{equation}\label{metric10d_for_reduction_S5}
ds^2_{10} = ds^2_5 + \frac{1}{g^2}D\mu^i D\mu^i \,,
\end{equation}
where $\mu^i\mu^i = 1$, $i\in\{1,2,3,4,5,6\}$, are the embedding coordinates of $S^5\subset\mb{R}^6$, $D\mu^i = d\mu^i + g A^{ij}_{(1)} \mu^j$, with only $A^{12}_{(1)}=A^{34}_{(1)}=A^{56}_{(1)}=\mc{A}_{(1)}$, and
\begin{equation}
F_{(5)} = (1+{\ast})G_{(5)} \,,
\end{equation}
with
\begin{equation}\label{flux10d}
G_{(5)} = 4g \vol_5 + \frac{1}{2g^2}{\ast}_5 F^{ij}_{(2)}\wedge D\mu^i\wedge D\mu^j \,,
\end{equation}
where $F^{ij}_{(2)} = dA^{ij}_{(1)} + gA^{ik}_{(1)}\wedge A^{kj}_{(1)}$. To facilitate the desired NATD transformations, we find it useful to choose the following the parameterisation of the embedding coordinates,
\begin{equation}\label{S5_parameters}
\mu^1 + i\mu^2 = \cos\xi\,\cos\frac{\theta}{2}\,e^{i(\psi+\phi)/2}\,,\quad \mu^3+i\mu^4= \cos\xi\,\sin\frac{\theta}{2}\,e^{i(\psi-\phi)/2} \,,\quad \mu^5+i\mu^6 = \sin\xi \,e^{i\tau} \,,
\end{equation}
with $\xi\in[0,\pi/2]$, $\tau\in[0,2\pi)$, $\theta\in[0,\pi]$, $\phi\in[0,2\pi)$, and $\psi\in[0,4\pi)$. Here we will use the same conventions for the $SU(2)$ isometry group as outlined in \cite{Itsios:2013wd}. The $SU(2)$ group element is parametrised by
\begin{equation}\label{su2groupelement}
\mathrm{g}=e^{\frac{i}{2}\phi \tau_3}\cdot e^{\frac{i}{2}\theta \tau_2}\cdot e^{\frac{i}{2}\psi \tau_3}\,,
\end{equation}
where $\tau_1,\tau_2,\tau_3$ are the standard Pauli matrices obeying $\tau_i\tau_j=\delta_{ij}\mathds{1}+i\epsilon_{ijk}\tau_k$, and the left-invariant Maurer-Cartan 1-forms are given by
\begin{equation}
\begin{split}\label{eq:SU(2)_mc_1_forms_S5}
L_1 &=\frac{1}{\sqrt{2}}\left( -\sin\psi\,d\theta + \sin\theta\cos\psi\,d\phi\right)\,,\\
 L_2 &=\frac{1}{\sqrt{2}}\left( \cos\psi\,d\theta + \sin\theta\sin\psi\,d\phi\right)\,,\\
  L_3 &=\frac{1}{\sqrt{2}}\left( d\psi + \cos\theta\,d\phi\right) \,,
\end{split}
\end{equation}
and they satisfy the algebra
\begin{equation}
dL_i =\frac{1}{\sqrt{2}}\epsilon_{ijk}L_j\wedge L_k \,.
\end{equation}
The ten-dimensional metric is now written as
\begin{equation}
\begin{split}
ds^2_{10} &= ds^2_5 + \frac{1}{g^2}\left[d\xi^2 + \sin^2\xi\left(d\tau - g\mc{A}_{(1)}\right)^2 \right.\\
&\left.\ph{ds^2_5 + \frac{1}{g^2}\quad}+ \frac{1}{2}\cos^2\xi\left(L_1^2+L_2^2 + \left[L_3-\sqrt{2}g\mc{A}_{(1)}\right]^2\right)\right] \,.
\end{split}
\end{equation}
Defining the orthonormal frame
\begin{equation}
\begin{split}\label{eq:pure_original_vielbeins_S5}
{e}^{\alpha}=&\,\bar{e}^{\alpha}\,,\quad \text{for ${\alpha}=0,1,2,3,4$}\\
{e}^5=&\,\frac{1}{g}\,d\xi\,,\quad {e}^6=\,\frac{1}{g}\sin\xi\left(d\tau - g\mc{A}_{(1)} \right)\,,\quad {e}^7=\,\frac{1}{\sqrt{2}g}\cos\xi\, L_1\,,\\
{e}^8=&\,\frac{1}{\sqrt{2}g}\cos\xi\, L_2\,,\quad {e}^9=\,\frac{1}{\sqrt{2}g}\cos\xi\, \left(L_3-\sqrt{2}g\mc{A}_{(1)}\right)\,,
\end{split}
\end{equation}
the 5-form flux is 
\begin{equation}
\begin{split}\label{eq:vielbein_F_5_on_S5}
F_{(5)}&= 4g \left(e^{01234}-e^{56789}\right)+\left( e^{78}\wedge{\ast_5\mc{F}_{(2)}}-e^{569} \wedge\mc{F}_{(2)}\right)\\
&\quad +\cos\xi \left(e^{56}\wedge{\ast_5\mc{F}_{(2)}}-e^{789}\wedge\mc{F}_{(2)} \right)-\sin\xi \left(e^{59}\wedge{\ast_5\mc{F}_{(2)}} + e^{678}\wedge\mc{F}_{(2)} \right)\,,
\end{split}
\end{equation}
where $\mc{F}_{(2)} =d\mc{A}_{(1)}$. The resulting theory of this truncation is the $D=5$ minimal gauged supergravity \cite{Gunaydin:1983bi}, and its action is given by
\begin{equation}
\mc{L}_{(5)} = R\vol_5 + 12g^2\vol_5-\frac{3}{2}\mc{F}_{(2)}\wedge{\ast_5}\mc{F}_{(2)} - \mc{F}_{(2)}\wedge \mc{F}_{(2)}\wedge \mc{A}_{(1)} \,.
\end{equation}
Setting $\mc{A}_{(1)} = \frac{2}{3}{A}_{(1)}$ to make contact with the conventions adopted in \cite{Ferrero:2020laf}, the equations of motion are 
\begin{equation}
\begin{split}
&R_{\mu\nu} = -4g^2g_{\mu\nu} + \frac{2}{3}{F}^{}_{\mu\rho}{F}_{\nu}^{\ph{\nu}\rho} - \frac{1}{9}g_{\mu\nu}\left({F}_{(2)}\right)^2 \,,\\
&d{\ast_5}{F}_{(2)} = -\frac{2}{3}{F}_{(2)}\wedge{F}_{(2)} \,,
\end{split}
\end{equation}
where ${F}_{(2)} =d{A}_{(1)}$.

The above discussion summarises the truncation of type \Romannum{2}B supergravity on $S^5$ to $D=5$ minimal gauged supergravity at the level of the bosonic fields. Now we turn our attention to the reduction of the IIB fermionic variations. In our truncation, we have $\Phi=C_{(0)}=B_{(2)}=C_{(2)}=0$, hence the condition for a bosonic background to preserve some supersymmetry is given by \cite{Schwarz:1983qr,Howe:1983sra}
\begin{equation}
\left[\nabla_{M}+\frac{i}{192}\Gamma^{N_1N_2N_3N_4}F_{MN_1N_2N_3N_4}\right]\epsilon=0\,,
\end{equation}
where the $\epsilon$ is a Weyl spinor with chirality $\Gamma_{11}\epsilon=-\epsilon$, where $\Gamma_{11} = \Gamma^0\cdots \Gamma^{9}$. Using the self-duality of $F_{(5)}$, this can be rewritten in a more convenient form,
\begin{equation}\label{eq:IIB_Killing_spinor_equation}
\left[\nabla_{M}+\frac{i}{16}\slashed{F}_{(5)}\Gamma_M\right]\epsilon=0\,,
\end{equation}
where 
\begin{equation}
\slashed{F}_{(5)} = \frac{1}{5!}F_{N_1\cdots N_5}\Gamma^{N_1\cdots N_5} \,.
\end{equation}
From \eqref{eq:vielbein_F_5_on_S5}, we find that
\begin{equation}
\begin{split}\label{eq:fey_contraction_five_form}
\slashed{F}_{(5)}&= 4g \left(\Gamma^{01234}-\Gamma^{56789}\right)+\frac{1}{12}\mc{F}_{\alpha\beta}\left(\epsilon^{\alpha\beta}_{\ph{\alpha\beta}\gamma_1\gamma_2\gamma_3}\Gamma^{\gamma_1\gamma_2\gamma_3}\Gamma^{78} - 6\Gamma^{\alpha\beta}\Gamma^{569} \right)\\
&\quad +\frac{1}{12}\cos\xi\,\mc{F}_{\alpha\beta}\left(\epsilon^{\alpha\beta}_{\ph{\alpha\beta}\gamma_1\gamma_2\gamma_3}\Gamma^{\gamma_1\gamma_2\gamma_3}\Gamma^{56}-6\Gamma^{\alpha\beta}\Gamma^{789} \right)\\
&\quad -\frac{1}{12}\sin\xi\,\mc{F}_{\alpha\beta} \left(\epsilon^{\alpha\beta}_{\ph{\alpha\beta}\gamma_1\gamma_2\gamma_3}\Gamma^{\gamma_1\gamma_2\gamma_3}\Gamma^{59}+6\Gamma^{\alpha\beta}\Gamma^{678}  \right)\,.
\end{split}
\end{equation}
Following \cite{Gauntlett:2005ww}, it is useful to decompose the Clifford algebra $\text{Cliff}(9,1)$ in the following way,
\begin{equation}
\begin{split}\label{eq:Clifford(9,1)_S5}
\Gamma^\alpha=&\rho^\alpha\otimes \mathds{1}\otimes \tau_3\,,\quad \text{for ${\alpha}=0,1,2,3,4$}\\
\Gamma^i=&\mathds{1}\otimes\gamma^i \otimes \tau_1\,,\quad \text{for $i=5,6,7,8,9$}
\end{split}
\end{equation}
with $\rho^\alpha$, the gamma matrices of $\text{Cliff}(4,1)$ satisfying $\rho^{01234}=-i\mathds{1}$ and $\gamma^i$, the gamma matrices of $\text{Cliff}(5,0)$ satisfying $\gamma^{56789}=\mathds{1}$. Hence we have
\begin{equation}
\Gamma_{11}=\mathds{1}\otimes\mathds{1}\otimes\tau_2 \,.
\end{equation}
Under this setup, we decompose the \Romannum{2}B spinor in the following way
\begin{equation}\label{iibspinordecomp}
\epsilon=\chi\otimes\lambda\otimes\varphi\,,
\end{equation}
such that $\tau_2\varphi=-\varphi$. Using \eqref{eq:fey_contraction_five_form} and taking the chirality condition into account, we find that
\begin{equation}
\begin{split}\label{eq:trick_formula}
\slashed{F}_{(5)}\Gamma_{\alpha}&=-8ig\rho_\alpha+i\mc{F}_{\beta\gamma}\rho^{\beta\gamma}\rho_{\alpha}\gamma^{78}+i\cos\xi\,\mc{F}_{\beta\gamma}\rho^{\beta\gamma}\rho_{\alpha}\gamma^{56}-i\sin\xi\,\mc{F}_{\beta\gamma}\rho^{\beta\gamma}\rho_{\alpha}\gamma^{59}\,,\\
\slashed{F}_{(5)}\Gamma_{i}&=-8g\gamma_i+\mc{F}_{\alpha\beta}\rho^{\alpha\beta}\gamma^{78}\gamma_i+\cos\xi\,\mc{F}_{\alpha\beta}\rho^{\alpha\beta}\gamma^{56}\gamma_i-\sin\xi\,\mc{F}_{\alpha\beta}\rho^{\alpha\beta}\gamma^{59}\gamma_i\,,
\end{split}
\end{equation}
where we have removed the tensor products for notational convenience. Combining the above results, we obtain
\begin{equation}
\begin{split}\label{eq:susy_variation_spacetime}
&\left[\nabla_{\alpha}^{(5)}+g\mc{A}_\alpha \gamma^{78}+\frac{i}{4}\sin\xi\,\mc{F}_{\alpha\beta}\rho^\beta\gamma^6+\frac{i}{4}\cos\xi\,\mc{F}_{\alpha\beta}\rho^\beta\gamma^9+\frac{g}{2}\rho_\alpha\right.\\
&\left.-\frac{1}{16}\mc{F}_{\beta\gamma}\rho^{\beta\gamma}\rho_{\alpha}\gamma^{78}-\frac{1}{16}\cos\xi\,\mc{F}_{\beta\gamma}\rho^{\beta\gamma}\rho_{\alpha}\gamma^{56}+\frac{1}{16}\sin\xi\,\mc{F}_{\beta\gamma}\rho^{\beta\gamma}\rho_{\alpha}\gamma^{59}\right.\\
&\left. + g\mc{A}_{\alpha}\partial_\tau + 2g\mc{A}_{\alpha}\partial_\psi \right]\epsilon=0\,,
\end{split}
\end{equation}
and
\begin{align*}\label{eq:susy_variation_on_S5_1}
&\left[\partial_5-\frac{ig}{2}\gamma_5+\frac{i}{16}\mc{F}_{\alpha\beta}\rho^{\alpha\beta}\gamma^{78}\gamma_5+\frac{i}{16}\cos\xi\,\mc{F}_{\alpha\beta}\rho^{\alpha\beta}\gamma^{56}\gamma_5-\frac{i}{16}\sin\xi\,\mc{F}_{\alpha\beta}\rho^{\alpha\beta}\gamma^{59}\gamma_5\right]\epsilon=0\,,\\
&\left[\partial_6+\frac{1}{8}\sin\xi\,\mc{F}_{\alpha\beta}\rho^{\alpha\beta}-\frac{g}{2}\cot\xi\gamma^{56}-\frac{ig}{2}\gamma_6\right.\\
&\left.+\frac{i}{16}\mc{F}_{\alpha\beta}\rho^{\alpha\beta}\gamma^{78}\gamma_6+\frac{i}{16}\cos\xi\,\mc{F}_{\alpha\beta}\rho^{\alpha\beta}\gamma^{56}\gamma_6-\frac{i}{16}\sin\xi\,\mc{F}_{\alpha\beta}\rho^{\alpha\beta}\gamma^{59}\gamma_6\right]\epsilon=0\,,\numberthis
\end{align*}
and
\begin{equation}
\begin{split}\label{eq:susy_variation_on_S5_2}
&\left[\partial_7+\frac{g}{2}\tan\xi\gamma^{57}+\frac{g}{2}\sec\xi\gamma^{89}-\frac{ig}{2}\gamma_7\right.\\
&\left.+\frac{i}{16}\mc{F}_{\alpha\beta}\rho^{\alpha\beta}\gamma^{78}\gamma_7+\frac{i}{16}\cos\xi\,\mc{F}_{\alpha\beta}\rho^{\alpha\beta}\gamma^{56}\gamma_7-\frac{i}{16}\sin\xi\,\mc{F}_{\alpha\beta}\rho^{\alpha\beta}\gamma^{59}\gamma_7\right]\epsilon=0\,,\\
&\left[\partial_8+\frac{g}{2}\tan\xi\gamma^{58}-\frac{g}{2}\sec\xi\gamma^{79}-\frac{ig}{2}\gamma_8\right.\\
&\left.+\frac{i}{16}\mc{F}_{\alpha\beta}\rho^{\alpha\beta}\gamma^{78}\gamma_8+\frac{i}{16}\cos\xi\,\mc{F}_{\alpha\beta}\rho^{\alpha\beta}\gamma^{56}\gamma_8-\frac{i}{16}\sin\xi\,\mc{F}_{\alpha\beta}\rho^{\alpha\beta}\gamma^{59}\gamma_8\right]\epsilon=0\,,\\
&\left[\partial_9+\frac{1}{8}\cos\xi\,\mc{F}_{\alpha\beta}\rho^{\alpha\beta}+\frac{g}{2}\tan\xi\gamma^{59}+\frac{g}{2}\sec\xi\gamma^{78}-\frac{ig}{2}\gamma_9\right.\\
&\left.+\frac{i}{16}\mc{F}_{\alpha\beta}\rho^{\alpha\beta}\gamma^{78}\gamma_9+\frac{i}{16}\cos\xi\,\mc{F}_{\alpha\beta}\rho^{\alpha\beta}\gamma^{56}\gamma_9-\frac{i}{16}\sin\xi\,\mc{F}_{\alpha\beta}\rho^{\alpha\beta}\gamma^{59}\gamma_9\right]\epsilon=0\,,
\end{split}
\end{equation}
where $\partial_{5,6,7,8,9}$ are the partial derivatives along the tangent space directions. The $5$ to $9$ components of the Killing spinor equation are solved by
\begin{equation}
\lambda = e^{-\frac{i\tau}{2}}e^{\frac{i\xi}{2}\gamma^5}\eta_0 \,,
\end{equation}
where the constant spinor $\eta_0$ obeys the projection conditions
\begin{equation}
\gamma^{56}\eta_0 = \gamma^{78}\eta_0 = -i\eta_0 \,.
\end{equation}
Substituting this into the $\alpha$ components of the Killing spinor equation, we find that it reduces to 
\begin{equation}\label{eq:KSE_S5_our_truncation}
\left[\nabla_{\alpha}^{(5)} -\frac{3ig}{2}\mc{A}_\alpha+\frac{i}{8}\left(\rho_{\alpha}^{\phantom{\alpha}\beta\gamma}-4\delta_{\alpha}^\beta\rho^\gamma\right)\mc{F}_{\beta\gamma}+\frac{g}{2}\rho_\alpha\right]\chi=0\,,
\end{equation}
which is the standard Killing spinor equation of the five-dimensional minimal gauged supergravity. Again, we can set $\mc{A}_{(1)} = \frac{2}{3}{A}_{(1)}$ to make contact with the conventions adopted in \cite{Ferrero:2020laf} and \eqref{eq:KSE_S5_our_truncation} becomes \eqref{eq:5d_minimal_KSE}.

\subsection{$T^{1,1}$ truncation}\label{sec:T11_truncation}

Using the results of \cite{Buchel:2006gb}, the $T^{1,1}$ truncation of type \Romannum{2}B supergravity to $D=5$ minimal gauged supergravity, at the level of the bosonic fields, is given by
\begin{align}\label{metric10d_for_reduction_T11}
\begin{split}
ds^2_{10} =& \,ds^2_5 + \frac{1}{g^2}\left[\frac{1}{6}\left(d\theta_1^2+\sin^2\theta_1\,d\phi_1^2\right)+\frac{1}{6}\left(d\theta_2^2+\sin^2\theta_2\,d\phi_2^2\right)\right.\\
&\left.\phantom{ds^2_5 + \frac{1}{g^2}[}+\frac{1}{9}\left(d\psi+\cos\theta_1\,d\phi_1+\cos\theta_2\,d\phi_2+g\hat{\mc{A}}_{(1)}\right)^2\right]\,,
\end{split}
\end{align}
and
\begin{equation}
F_{(5)} = (1+{\ast})\left[4g\text{vol}_5+\frac{1}{18g^2}\left(\sin\theta_1\,d\theta_1\wedge d\phi_1+\sin\theta_2\,d\theta_2\wedge d\phi_2\right)\wedge{\ast_5 \hat{\mc{F}}_{(2)}}\right] \,.
\end{equation}
To facilitate the desired NATD transformations, it is useful to introduce the following left-invariant Maurer-Cartan 1-forms for the $S^3$,
\begin{equation}
\begin{split}\label{eq:mc_1_form_T11}
\sigma_7&=\cos\psi \sin\theta_2\,d\phi_2-\sin\psi\,d\theta_2\,, \\
 \sigma_8&=\sin\psi \sin\theta_2\,d\phi_2+\cos\psi\,d\theta_2\,,\\
\sigma_9&=d\psi+\cos\theta_2\, d\phi_2\,.
\end{split}
\end{equation}
We will relate these to the Maurer-Cartan 1-forms defined in \eqref{eq:SU(2)_mc_1_forms_S5} by $\sigma_{7,8,9} =\sqrt{2}L_{1,2,3}$. 

This allows us to recast the ten-dimensional metric as
\begin{align}
\begin{split}
ds^2_{10} =&\, ds^2_5 + \frac{1}{g^2}\left[\frac{1}{6}\left(d\theta_1^2+\sin^2\theta_1\,d\phi_1^2\right)+\frac{1}{3}\left(L_1^2+L_2^2\right)\right.\\
&\left.\phantom{ds^2_5 + \frac{1}{g^2}[}+\frac{1}{9}\left(\sqrt{2}L_3+\cos\theta_1\,d\phi_1+g\hat{\mc{A}}_{(1)}\right)^2\right]\,.
\end{split}
\end{align}
We define the ten-dimensional orthonormal frame as follows
\begin{equation}
\begin{split}\label{eq:pure_original_vielbeins_T11}
{e}^{\alpha}&=\,\bar{e}^{\alpha}\,,\quad \text{for ${\alpha}=0,1,2,3,4$}\\
{e}^5&=\,\frac{1}{\sqrt{6}g}\sin\theta_1\,d\phi_1\,\,,\quad {e}^6=\,\frac{1}{\sqrt{6}g}d\theta_1\,,\quad {e}^7=\,\frac{1}{\sqrt{3}g}\,L_1\,,\\
 {e}^8&=\,\frac{1}{\sqrt{3}g}\,L_2\,,\quad {e}^9=\,\frac{1}{3g}\left(\sqrt{2}L_3+\cos\theta_1\,d\phi_1+g\hat{\mc{A}}_{(1)}\right)\,,
\end{split}
\end{equation}
where $\bar{e}^\alpha$ define the orthonormal frame of $ds_5^2$. The self-dual five-form flux can now be written as
\begin{equation}\label{eq:vielbein_F_5_on_T11}
F_{(5)}= 4g \left(e^{01234}-e^{56789}\right)-\frac{1}{3}\left(e^{56}+e^{78}\right)\wedge {\ast_5 \hat{\mc{F}}_{(2)}}+\frac{1}{3}\left(e^{569}+e^{789}\right)\wedge {\hat{\mc{F}}_{(2)}}\,.
\end{equation}
The resulting theory of this truncation is again the 5-dimensional minimal gauged supergravity \cite{Gunaydin:1983bi}, and its action is given by
\begin{equation}
\mc{L}_{(5)} = R\vol_5 + 12g^2\vol_5-\frac{1}{6}\hat{\mc{F}}_{(2)}\wedge{\ast_5}\hat{\mc{F}}_{(2)} + \frac{1}{27} \hat{\mc{F}}_{(2)}\wedge \hat{\mc{F}}_{(2)}\wedge \hat{\mc{A}}_{(1)} \,.
\end{equation}
Setting $\hat{\mc{A}}_{(1)} = -2{A}_{(1)}$ to make contact with the conventions adopted in \cite{Ferrero:2020laf}, the equations of motion are back to
\begin{equation}
\begin{split}
&R_{\mu\nu} = -4g^2g_{\mu\nu} + \frac{2}{3}{F}^{}_{\mu\rho}{F}_{\nu}^{\ph{\nu}\rho} - \frac{1}{9}g_{\mu\nu}\left({F}_{(2)}\right)^2 \,,\\
&d{\ast_5}{F}_{(2)} = -\frac{2}{3}{F}_{(2)}\wedge{F}_{(2)} \,,
\end{split}
\end{equation}
where ${F}_{(2)} =d{A}_{(1)}$.

The above discussion summarises the truncation of type \Romannum{2}B supergravity on $T^{1,1}$ to $D=5$ minimal gauged supergravity at the level of the bosonic fields. Now we turn our attention to the reduction of the IIB fermionic variations. We will use the same Clifford algebra decomposition as in \eqref{eq:Clifford(9,1)_S5}, and the same spinor decomposition \eqref{iibspinordecomp}.
Repeating the same procedure as we have for the $S^5$ reduction, we find that the Killing spinor equations along the $T^{1,1}$ directions are solved with $\lambda = \eta_0$ being a constant spinor satisfying
\begin{equation}
\gamma^{56}\eta_0 = \gamma^{78}\eta_0 = -i\eta_0\,.
\end{equation}
This is in contrast with the results of \cite{Buchel:2006gb}, where their Killing spinor obeys both the projection conditions above, as well as having a functional dependence on the $SU(2)$ directions given by a factor of $e^{i\psi/2}$. The reason for this apparent disagreement is that the orthonormal frame used in \cite{Buchel:2006gb} is different from the one that we are using \eqref{eq:pure_original_vielbeins_T11}. These two frames are related by a local Lorentz transformation, under which a spinor transforms linearly. Going from our frame to the frame of \cite{Buchel:2006gb}, the linear transformation of the spinor induces precisely the extra factor of $e^{i\psi/2}$. The orthonormal frame we chose here is manifestly $SU(2)$ invariant, and as a consequence, the Killing spinor is also $SU(2)$ invariant. 

With the projection conditions, the Killing spinor equations along the worldvolume then reduce to \eqref{eq:5d_minimal_KSE}
\begin{equation}
\left[\nabla_{\alpha}^{(5)} - igA_\alpha + \frac{i}{12}\left(\rho_{\alpha}^{\phantom{\alpha}\beta\gamma}-4\delta_{\alpha}^\beta\rho^\gamma\right)F_{\beta\gamma}+\frac{g}{2}\rho_\alpha\right]\chi=0\,.
\end{equation}

\subsection{$Y^{p,q}$ truncation}\label{sec:Ypq_truncation}
Using the results of \cite{Buchel:2006gb}, the $Y^{p,q}$ truncation of type \Romannum{2}B supergravity to $D=5$ minimal gauged supergravity, at the level of the bosonic fields, is similar to that for $T^{1,1}.$ The ten-dimensional metric is given by\footnote{We have taken $\phi\rightarrow-\phi$ in the metric, when compared to \cite{Buchel:2006gb,Gauntlett:2004yd,Martelli:2004wu}, to make contact with the conventions used in \cite{Sfetsos:2014tza}.}
\begin{align}\label{metric10d_for_reduction_Ypq}
\begin{split}
ds^2_{10} =&\, ds^2_5 + \frac{1}{g^2}\left[\frac{1-\tilde{y}}{6}\left(d\theta^2+\sin^2\theta\,d\phi^2\right)+\frac{w(\tilde{y})v(\tilde{y})}{36}\left(d\beta-\cos\theta\,d\phi\right)^2\right.\\
&\left.+\frac{1}{w(\tilde{y})v(\tilde{y})}d\tilde{y}^2+\frac{1}{9}\left(d\psi+\cos\theta\,d\phi+\tilde{y}\left(d\beta-\cos\theta\,d\phi\right)+g\tilde{\mc{A}}_{(1)}\right)^2\right]\,,
\end{split}
\end{align}
where the functions are defined by
\begin{align}
w(\tilde{y})=\frac{2(b-\tilde{y}^2)}{1-\tilde{y}}\,,\quad  v(\tilde{y})=\frac{b-3\tilde{y}^2+2\tilde{y}^3}{b-\tilde{y}^2}\,,
\end{align}
with parameter $0<b<1$ which is determined by $p$ and $q$. The five-form flux is given by
\begin{equation}
F_{(5)} = (1+{\ast})\left[4g\text{vol}_5-\frac{1}{3g^2}J_{(2)}\wedge{\ast_5 \tilde{\mc{F}}_{(2)}}\right] \,,
\end{equation}
where the K\"{a}hler 2-form $J_{(2)}$ is
\begin{equation}
J_{(2)}=-\frac{1-\tilde{y}}{6}\sin\theta\,d\theta\wedge d\phi+\frac{1}{6}d\tilde{y}\wedge \left(d\beta-\cos\theta\,d\phi\right)\,,
\end{equation}
To facilitate the desired NATD transformations, it is useful to redefine $\beta=-6\alpha-\psi$, which allows us to recast the ten-dimensional metric as
\begin{equation}
\begin{split}
ds^2_{10} =&\, ds^2_5 + \frac{1}{g^2}\left[\frac{1-\tilde{y}}{3}\left(L_1^2+L_2^2\right)+\frac{1}{w(\tilde{y})v(\tilde{y})}d\tilde{y}^2+k(\tilde{y})^2\left(d\alpha-\frac{g}{6}\tilde{\mc{A}}_{(1)}\right)^2\right.\\
&\left.\phantom{ds^2_5 + \frac{1}{g^2}[}+\left(\sqrt{2\Delta(\tilde{y})}\,L_3+h(\tilde{y})d\alpha+\frac{g(1-\tilde{y})}{9\sqrt{\Delta(\tilde{y})}}\tilde{\mc{A}}_{(1)}\right)^2\right]\,,
\end{split}
\end{equation}
where $L_{1,2,3}$ are the Maurer-Cartan 1-forms given in \eqref{eq:SU(2)_mc_1_forms_S5}, and 
\begin{equation}
\begin{split}
f(\tilde{y})&=\frac{b-2\tilde y+\tilde y^2}{6(b-\tilde y^2)}\,,\quad \Delta(\tilde{y})=\frac{v(\tilde{y})}{9}+w(\tilde{y})f(\tilde{y})^2\,,\\
h(\tilde{y})&=\frac{w(\tilde{y})f(\tilde{y})}{\sqrt{\Delta(\tilde{y})}}\,,\quad k(\tilde{y})^2=\frac{v(\tilde{y})w(\tilde{y})}{9\Delta(\tilde{y})}\,.
\end{split}
\end{equation}
We define the ten-dimensional orthonormal frame as follows
\begin{equation}
\begin{split}\label{eq:pure_original_vielbeins_Ypq}
{e}^{\alpha}=&\,\bar{e}^{\alpha}\,,\quad \text{for ${\alpha}=0,1,2,3,4$}\\
{e}^5=&\,\frac{1}{g\sqrt{wv}}\,d\tilde{y}\,,\quad {e}^6=\,\frac{k}{g}\,\left(d\alpha-\frac{g}{6}\tilde{\mc{A}}_{(1)}\right)\,,\quad  {e}^7=\,\frac{\sqrt{1-\tilde{y}}}{\sqrt{3}g}\,L_1\,,\\
{e}^8=&\,\frac{\sqrt{1-\tilde{y}}}{\sqrt{3}g}\,L_2\,,\quad {e}^9=\,\frac{1}{g}\left(\sqrt{2\Delta}\,L_3+h\,d\alpha+\frac{g(1-\tilde{y})}{9\sqrt{\Delta}}\tilde{\mc{A}}_{(1)}\right)\,,
\end{split}
\end{equation}
where $\bar{e}^\alpha$ define the orthonormal frame of $ds_5^2$. The self-dual five-form flux can now be written as
\begin{align*}\label{eq:vielbein_F_5_on_Ypq}
F_{(5)}&= 4g \left(e^{01234}-e^{56789}\right) -\frac{1}{3}\left(e^{78}-\frac{3}{\sqrt{\Delta}}\left(\Delta-\frac{wf}{6}\right)e^{56} - \frac{1}{6}\sqrt{\frac{wv}{\Delta}}e^{59}\right)\wedge {{\ast}_5\tilde{\mc{F}}_{(2)}} \\
&\quad+\frac{1}{3}\left(e^{569}-\frac{3}{\sqrt{\Delta}}\left(\Delta-\frac{wf}{6}\right)e^{789} + \frac{1}{6}\sqrt{\frac{wv}{\Delta}}e^{678}\right)\wedge \tilde{\mc{F}}_{(2)} 
\,.\numberthis
\end{align*}
The resulting theory of this truncation is again the 5-dimensional minimal gauged supergravity \cite{Gunaydin:1983bi}, and its action is given by
\begin{equation}
\mc{L}_{(5)} = R\vol_5 + 12g^2\vol_5-\frac{1}{6}\tilde{\mc{F}}_{(2)}\wedge{\ast_5}\tilde{\mc{F}}_{(2)} -\frac{1}{27} \tilde{\mc{F}}_{(2)}\wedge \tilde{\mc{F}}_{(2)}\wedge \tilde{\mc{A}}_{(1)} \,.
\end{equation}
Setting $\tilde{\mc{A}}_{(1)} = 2{A}_{(1)}$ to make contact with the conventions adopted in \cite{Ferrero:2020laf}, the equations of motion are back to
\begin{equation}
\begin{split}
&R_{\mu\nu} = -4g^2g_{\mu\nu} + \frac{2}{3}{F}^{}_{\mu\rho}{F}_{\nu}^{\ph{\nu}\rho} - \frac{1}{9}g_{\mu\nu}\left({F}_{(2)}\right)^2 \,,\\
&d{\ast_5}{F}_{(2)} = -\frac{2}{3}{F}_{(2)}\wedge{F}_{(2)} \,,
\end{split}
\end{equation}
where ${F}_{(2)} =d{A}_{(1)}$.

The above discussion summarises the truncation of type \Romannum{2}B supergravity on $Y^{p,q}$ to $D=5$ minimal gauged supergravity at the level of the bosonic fields. Now we turn our attention to the reduction of the IIB fermionic variations. We will use the same Clifford algebra decomposition as in \eqref{eq:Clifford(9,1)_S5}, and the same spinor decomposition \eqref{iibspinordecomp}. Since the computation is rather more involved than the $T^{1,1}$ case, we will record here some useful results.
The non-zero components of the spin connection are
\begin{equation}
\begin{split}
&\omega_{\alpha\beta}=\bar{\omega}_{\alpha\beta}+  \frac{k}{6}F_{\alpha\beta} e^6-\frac{1-\tilde y }{9\sqrt{\Delta}}F_{\alpha\beta}e^9\,,\quad \omega_{\alpha6}= \frac{k}{6}F_{\alpha\beta} e^\beta\,,\quad \omega_{\alpha9}=-\frac{1-\tilde y }{9\sqrt{\Delta}}F_{\alpha\beta}e^\beta\,,\\
&\omega_{56}=- \frac{3g(1-\tilde y)kh}{wv}e^6+\frac{g (1-b)}{9\Delta(1-\tilde y)}e^9\,,\quad \omega_{57}=\frac{g\sqrt{wv}}{2(1-\tilde y)} e^7\,,\quad \omega_{58}=\frac{g\sqrt{wv}}{2(1-\tilde y)} e^8\,,\\
&\omega_{59}=\frac{g (1-b)}{9\Delta(1-\tilde y)}e^6- \frac{3gk}{2(1-\tilde y)\sqrt{\Delta}}\left(\Delta-\frac{1-\tilde y}{3}\right)e^9\,,\quad \omega_{69}=- \frac{g (1-b)}{9\Delta(1-\tilde y)}e^5\,,\\
&\omega_{78}=-\frac{gh}{k\sqrt{\Delta}}e^{6} -2gA_\alpha e^{\alpha} + \frac{g}{\sqrt{\Delta}}\left(1-\frac{3\Delta}{1-\tilde y}\right)e^{9}\,,\\
&\omega_{79} = -\frac{3g\sqrt{\Delta}}{1-\tilde y}e^8\,,\quad \omega_{89}= \frac{3g\sqrt{\Delta}}{1-\tilde y}e^7\,,
\end{split}
\end{equation}
where we have set $\tilde{\mc{A}}_{(1)} = 2A_{(1)}$, and the flux contractions are 
\begin{equation}
\begin{split}
&\slashed{F}_{(5)}\Gamma_\alpha = -8ig\rho_\alpha - \frac{2i}{3}F_{\beta\gamma}\rho^{\beta\gamma}\rho_\alpha\left(\gamma_{78}-\frac{3}{\sqrt{\Delta}}\left(\Delta-\frac{wf}{6}\right)\gamma_{56} - \frac{1}{6}\sqrt{\frac{wv}{\Delta}}\gamma_{59}\right) \,,\\
&\slashed{F}_{(5)}\Gamma_i = -8g\gamma_i - \frac{2}{3}F_{\alpha\beta}\rho^{\alpha\beta}\left(\gamma_{78}-\frac{3}{\sqrt{\Delta}}\left(\Delta-\frac{wf}{6}\right)\gamma_{56} - \frac{1}{6}\sqrt{\frac{wv}{\Delta}}\gamma_{59}\right)\gamma_i \,.
\end{split}
\end{equation}
The $M = 5$ component of the Killing spinor equation \eqref{eq:IIB_Killing_spinor_equation} is then given by
\begin{equation}
\begin{split}
&\left[g\sqrt{wv}\partial_{\tilde y} + \frac{g(1-b)}{18\Delta(1-\tilde y)}\gamma_{578} - \frac{ig}{2}\gamma_5 \right.\\
&\left.- \frac{i}{24}F_{\alpha\beta}\rho^{\alpha\beta}\left(\gamma_{578}+\frac{3}{\sqrt{\Delta}}\left(\Delta-\frac{wf}{6}\right)\gamma_{6} + \frac{1}{6}\sqrt{\frac{wv}{\Delta}}\gamma_{5678}\right)\right]\lambda = 0\,.
\end{split}
\end{equation}
The solution to this is 
\begin{equation}\label{ypqKSElambda}
\lambda = e^{\frac{i}{2}L(\tilde y)\gamma_5}\eta \,, 
\end{equation}
where the $\tilde y$-independent spinor $\eta$ obeys the projection conditions
\begin{equation}\label{ypqproj}
\gamma^{56}\eta = -i\eta \,,\quad \gamma^{78} \eta = i\eta \,,
\end{equation}
and the function $L(\tilde y)$ is defined by
\begin{equation}
\cot L(\tilde y) = \frac{2(1-\tilde y)}{\sqrt{wv}} \,.
\end{equation}
For the remaining components of the Killing spinor equation along the $Y^{p,q}$ directions, we find, after some algebra, that they vanish identically if $\eta=\eta_0$ is a constant spinor along these directions. Substituting \eqref{ypqKSElambda} into the worldvolume components of the Killing spinor equation, and taking care to impose the projection conditions \eqref{ypqproj}, we find that it reduces to
\begin{equation}
\left[\nabla_{\alpha}^{(5)} -igA_\alpha+\frac{i}{12}\left(\rho_{\alpha}^{\phantom{\alpha}\beta\gamma}-4\delta_{\alpha}^\beta\rho^\gamma\right)F_{\beta\gamma}+\frac{g}{2}\rho_\alpha\right]\chi=0\,,
\end{equation}
which is the Killing spinor equation for the minimal gauged supergravity in five dimensions \eqref{eq:5d_minimal_KSE}.

\end{appendix}

\end{document}